\documentstyle[nato]{crckapb}      
\input epsf

\def\VEV#1{\left\langle #1\right\rangle}
\def\lt     {<}	
\def\gt     {>}	
\def\lsim   {_\sim^<}    
\def\gsim   {_\sim^>}    
\def\etal   {{\rm et~al.}}
\def\aj     {Ap. J.}
\def\araa   {Ann. Rev. Astron. \& Astrophysics}
\def\aap    {Astronomy \& Astrophysics}
\def\pr     {Phy. Rev.}

\def\ppnp   {ppnp}

\begin{opening}
\title{THE CMB ANISOTROPY EXPERIMENTS}
\subtitle{Cosmic Microwave Background}

\author{George F. Smoot}
\institute{Lawrence Berkeley National Lab \& Physics Department\\
           University of California \\ Berkeley CA 94720}

\end{opening}

\runningtitle{THE CMB ANISOTROPY}

\begin{document}


\section{Abstract}
Anisotropies in the cosmic microwave background (CMB) encode information
about the evolution and development of the Universe. 
It is understood that quality observations of the CMB anisotropies
can provide a very strong test of cosmological models and
provide high precision measurements of major cosmological parameters.
This paper provides a review of the COBE DMR results,
the current status of the measurements of the CMB anisotropy power spectrum
and then focuses on the programs that are likely to provide additional
results including both suborbital observations and the two selected
satellite missions: the NASA MIDEX mission MAP and the ESA M3 mission 
Max Planck Surveyor (formerly COBRAS/SAMBA).
This review includes both a description of the experimental programs and the
expected quality level of results. 
 
\section{Introduction}
The observed cosmic microwave background (CMB) radiation provides
strong evidence for the big bang model of cosmology
and is the best probe we have for determining conditions in the early Universe 
as well as determining many important cosmological parameters.
The angular power spectrum of the CMB contains information
on virtually all cosmological parameters of interest,
including the geometry of the Universe ($\Omega$),
the baryon density ($\Omega_B$), the Hubble expansion rate ($h$),
the cosmological constant ($\Lambda$), the number of light neutrinos ($n_\nu$),
the ionization history of the Universe, and the amplitudes
and spectral indices of the primordial and tensor perturbation spectra.
Precise CMB observations, data analysis, and interpretation can
distinguish between cosmological models. They can be used
to verify that the range of models under consideration is plausible
and to distinguish between models with primordial perturbations 
(e.g. the inflationary big bang)
or those with active perturbations (e.g. topological defects which
must result from spontaneous symmetry breaking of unified forces).
Once a model is thus singled out, its parameters can, in principle,
be determined to accuracies of the order of a per cent\cite{Jungman96}.

Since the initial detection of CMB temperature anisotropies by
the {\it COBE} DMR \cite{Smoot92}, over a dozen other balloon-borne
and ground-based experiments have reported anisotropy detections on
smaller angular scales.
With the existence of anisotropies now firmly established, 
observational goals have shifted towards an accurate determination
of the CMB anisotropy power spectrum over a wide range of angular scales.
The reasons for this are two-fold. (1) If the processes producing the
initial fluctuations are stochastic and random phase, then the power
spectrum contains all the information of the underlying physical model.
(2) It is observationally easier to obtain a power spectrum than a
fully reliable map.
Several technical advances, including improved receivers,
interferometry, and long-duration balloon flights, hold great promise
for high-precision maps in the next few years.
Ultimately, the two approved satellites: NASA MidEX mission MAP and
ESA M3 mission Planck are expected to provide high-angular-resolution
high-sensitivity maps of the entire sky in multifrequency bands.
Thus we can anticipate increasingly complex data sets requiring
sophisticated analysis: COBE DMR 4-year maps (6144 pixels),
the CfPA balloon experiments MAXIMA/BOOMERANG (26,000 to 130,000 pixels),
MAP ($\approx$400,000 pixels), and Planck ($> 10^6$ pixels).
These maps then hold the promise of revolutionizing cosmology
in terms of making it significantly more precise in quantitative terms.

It is the goal of these lectures to provide 
the background necessary to understand the existing data, soon to be
achieved data from experiments in progress, and finally the forthcoming
data from the more advanced experiments and space missions.
We proceed with some historical context, a review of the COBE observations
both for the discovery of anisotropy and as a prototype
for the next generation of space missions, a review of 
the current and proposed generations of balloon-borne experiments and 
interferometers, and finally a discussion of the new space-based experiments.

\section{CMB Background}
Primordial nucleosynthesis calculations require a cosmic background radiation (CBR)
with a temperature $kT \sim 1$~MeV at a redshift of $z \sim 10^9$.
Gamow, Alpher, \& Herman \cite{Gamow48} realized that this CBR was required
and predicted its evolution to a faint residual relic radiation
with a current temperature of a few degrees.  
Our more modern view of the hot big bang models gives the cosmic background
radiation a very central role in the development of the Universe.

The CMB was serendipitously discovered by Penzias \& Wilson \cite{Penzias65} 
in 1964 (published in 1965) 
and they noted that it was isotropic 
to the sensitivity of their measurement ($< 10$\%).
The observed CMB spectrum is well characterized by a $2.73 \pm 0.01$~K
blackbody spectrum. The hot big bang model predicts that the CBR
should have a thermal spectrum and this is verified precisely.
Combined with the observed spectrum of the dipole anisotropy,
this precision thermal spectrum also provides us with the knowledge and ability
to separate CMB anisotropies from the various foregrounds.
Anisotropies will in general have a spectrum set by the derivative 
of the CMB spectrum. See my previous lectures for a more complete discussion
of the expected anisotropy spectrum.

\section{Theoretical Anisotropies}
There are three primary threads of science that
are pursued by CMB anisotropy measurements:

(1) Initial Conditions for Large Scale Structure Formation:
The formation of galaxies, clusters of galaxies, and large scale
structures is a key issue in cosmology. 
Theory indicates that whatever the seeds of structure formation,
they will leave their imprint as anisotropies in the CMB.
Different scenarios for structure formation will leave different anisotropies.

(2) Physics of the Early Universe:
CMB anisotropy measurements are a probe of the ultra-high energy physics
and processes that occur in the very early universe.
These observations are a probe of inflation or quantum gravity 
and a test of potential topological defects (monopoles, strings, domain walls, 
and textures) that must result from spontaneous symmetry breaking.

(3) Geometry and Dynamics of the Universe:
Observations of the CMB anisotropies provide information on 
the metric and topology of the space-time,
the isotropy of expansion of space-time,
the curvature of space, and the possible rotation and shear 
of the Universe.

There is now a fairly extensive literature on the first two areas
and the third is what we consider classical cosmology.
In the early history of the field all measurements led only to
upper limits on CMB anisotropy and these in turn led to limits on 
but not a measurement of these processes.

A major finding of the initial {\it COBE} DMR discovery \cite{Smoot92}
was that the CMB was anisotropic on all observed angular scales.
A key question is what these anisotropies represent.
Immediately, the interpretation focused upon the seeds of large
scale structure formation.
In the early 1970's the observed large scale structure and scaling arguments 
led to the prediction \cite{Peebles70},\cite{Harrison70},\cite{Zeldovich72} 
that the primordial gravitational potential perturbations must have an equal
$rms$ amplitude on all scales.
This corresponds to a matter density perturbation power-law spectrum, 
$P(k)\propto k^n$,
where $k$ is the comoving wavenumber, with $n=1$.  
At that time there were no known mechanisms for producing
such a scale-invariant power spectrum of fluctuations.
In 1982 it was found that inflationary models predicted nearly 
scale-invariant perturbations as a result of quantum mechanical fluctuations 
at very early times.
Even with the proliferation of inflationary models, it is found that 
essentially all reasonable inflationary models predict $n \approx 1$. 
Presumably, a more reasonable class of inflationary models
will result in requiring a tie to particle physics.
It is now known that topological defects naturally produce scale-invariant 
fluctuations.
Thus there are at least two known mechanisms for producing a nearly
scale invariant primordial perturbation spectrum.

The translation from a scale-invariant spectrum of perturbations
to the CMB temperature anisotropies depends upon angular scale
and the contents of the universe. 
On large angular scales the results of most models are fairly similar.
Including the effects of a standard cold dark matter model, 
a Harrison-Zeldovich $n=1$ universe 
is consistent with the power spectrum measured by 
the {\it COBE} DMR data. 
The observed power spectrum of fluctuation amplitudes is also consistent 
with models of large scale structure formation based upon primordial seeds 
produced by quantum fluctuations or topological defects
in the early universe.

The physics of anisotropy caused by primordial density perturbations
is usually divided into four generic areas although they are all
treated properly in the full Sachs-Wolfe effect \cite{Sachs67}.
These effects are: 
the gravitational redshift which dominates at large angular scales,
the (Rees-Sciama) effect on light propagating through a changing potential,
the Doppler effect caused by the motion of the observer or the source, and
temperature or entropy variations.

Models of the formation of structure in the Universe fall into two broad 
classes: inflationary and defect models. 
Each model predicts an angular power spectrum of CMB anisotropy
described in terms of the amplitude of the spherical harmonic 
of multipole order $\ell$.

The most detailed theoretical work has been carried out for inflationary models.
Hu and White \cite{Hu96} have argued that all inflationary models
produce an angular power spectrum with a unique set of ``doppler" or
``acoustic" peaks between $\ell = 100$ and $\ell = 1000$ 
($11^\prime < \theta < 1.8^\circ$).
The relative position and height and the detailed shape of the peaks
provides more independent constraints than there are parameters
in the inflationary scenario and allow their determination.

Defect models, including cosmic strings and textures, provide
an alternative to inflation. Such models predict a non-gaussian distribution
of temperature fluctuations and a power spectrum different from 
that of inflation.

Most cosmological models do not predict the exact CMB temperature pattern 
that would be observed in our sky, but rather predict a statistical 
distribution of anisotropies. 
In the context of such models, the CMB temperature observed
in our sky is only a single realization drawn from 
the cosmic statistical distribution.
Theoretical models most often predict a power spectrum
in spherical harmonic amplitudes; 
as the physics of the models leads to primordial fluctuations
that are Gaussian random fields, the power spectrum is
sufficient to characterize the results.
Observations of the sky can be expressed as a
spherical harmonic temperature expansion
$T(\theta,\phi)=\sum_{\ell m} a_{\ell m}Y_{\ell m}(\theta,\phi)$.
If the original perturbations are Gaussian random fields,
the $a_{\ell m}$ are Gaussianly distributed,
and the power at each $\ell$ is $(2 \ell +1) C_\ell/(4\pi)$,
where $C_\ell \equiv \VEV{|a_{\ell m}|^2}$, 
is sufficient to characterize the results.
For an idealized full-sky observation, the variance of
each measured $C_\ell$ is $[2 /(2 \ell +1 )] C^2_\ell$.
This sampling variance (known as cosmic variance) comes about because
each $C_\ell$ is chi-squared distributed with $( 2 \ell +1 )$ degrees
of freedom for our observable volume of the Universe \cite{White94}.
Thus, in addition to experimental uncertainties, we account for the
{\it cosmic sample variance} uncertainties due to our observation
of a single realization in our analyses of the DMR maps.
Cosmic variance exists independently of the quality of the experiment.  
The power spectrum from the 4-year DMR map is cosmic variance limited for
$\ell \lsim 20$.

\section{The Legacy of COBE} 

The Cosmic Background Explorer ({\it COBE}) was NASA's first satellite
dedicated to cosmology and is a milestone for that and for the legacy
of information that it has provided on the early Universe.
Much of its results are from observations of the cosmic microwave background 
(CMB). The CMB is a pillar of the Big Bang model and encodes information about 
energy release in the early Universe, primordial perturbations, 
and the geometry of the universe.
As mentioned previously the CMB contains information on critical cosmological 
parameters such as $\Omega_0$, $\Omega_B$, $\Omega_\Lambda$, and $H_0$.
{\it COBE's} legacy of the precise measurement of the CMB spectrum and 
the discovery and early mapping of the CMB anisotropy low-$\ell$ power spectrum
provides a position from which to carry out a program testing our cosmological 
theories and understanding the early Universe precisely.

The Differential Microwave Radiometers (DMR) experiment
(\cite{Smoot90}) 
discovered 
CMB anisotropies from analysis of its first year of data
\cite{Smoot92}, \cite{Bennett92}, \cite{Wright92}, \cite{Kogut92}.
The CMB temperature fluctuations were measured at an angular
resolution of $7^\circ$ at frequencies of 31.5, 53, and 90 GHz.  
These results were supported by a detailed examination of the DMR calibration 
and its uncertainties (\cite{Bennett91}) and a detailed treatment of the upper
limits on residual systematic errors (\cite{Kogut92}).  
The {\it COBE} results were confirmed by the positive
cross-correlation between the {\it COBE} data and data from balloon-borne
observations at a shorter wavelength \cite{Ganga93}
and later by comparison of the {\it COBE} data and data from the ground-based
Tenerife experiment \cite{Lineweaver95} at longer wavelengths.
The positive correlation at both longer and shorter wavelengths
provides confidence in the results.
The results from analysis of two years of DMR data \cite{Bennett94} 
reconfirmed the results from the first year data.

This section summarizes the key and most recent results from COBE. 
Details can be found in the original references and in the most recent
FIRAS paper \cite{Fixsen96} and in a set of DMR 4-year analysis papers 
\cite{Banday96a}, \cite{Banday96b}, \cite{Bennett96}, 
\cite{Gorski96}, \cite{Hinshaw96a}, \cite{Hinshaw96b}, 
\cite{Lineweaver96}, \cite{Kogut96a}, \cite{Kogut96b}, \cite{Kogut96c}, 
\cite{Wright96}.

\subsection{The {\it COBE} DMR Instruments \& Data Analysis}
\begin{figure}
\centerline{\epsfxsize= 7 cm \epsfbox[0 0 478 684]{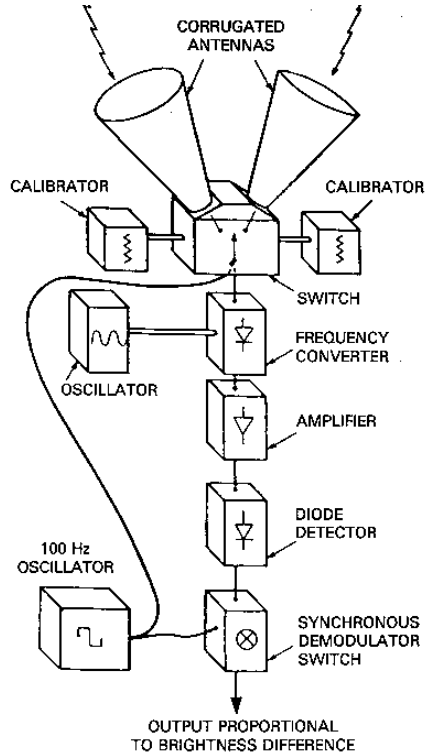}}
  \caption
{DMR signal flow schematic
}
\label{DMRsc}
\end{figure}

The DMR consists of 6 differential microwave radiometers: 2 nearly independent
channels, labeled A and B, at frequencies 31.5, 53, and 90 GHz (wavelength
9.5, 5.7, and 3.3 mm respectively). 
Each radiometer measures the difference in power between two 7$^\circ$ fields 
of view separated by 60$^\circ$, 30$^\circ$ to either side
of the spacecraft spin axis \cite{Smoot90}.  
Figure \ref{DMRsc} shows a schematic signal path for the DMRs.
{\it COBE} was launched from Vandenberg Air Force Base on 18 November 1989 
into a 900 km, 99$^\circ$ inclination circular orbit, 
which precesses to follow the terminator (light dark line on the Earth)
as the Earth orbits the Sun.
Attitude control keeps the spacecraft pointed away from the Earth and nearly
perpendicular to the Sun with a slight backward tilt so that solar radiation 
never directly illuminates the aperture plane.  
The combined motions of the spacecraft spin (75 s period),
orbit (103 m period), and orbital precession ($\sim 1^\circ$ per day) allow
each sky position to be compared to all others through a highly redundant set
of temperature difference measurements spaced 60$^\circ$ apart.
The on-board processor box-car integrates the differential signal 
from each channel for 0.5 s, and records the digitized differences 
for daily playback to a ground station.

Ground data processing consists of calibration, extensive systematic error
analyses, and conversion of time-ordered-data to sky maps \cite{Kogut96a}.
Checks on the correlated noise in the maps \cite{Lineweaver94} 
due to the map-making process indicate they are well below the 1\%\ level. 
The DMR time-ordered-data include systematic effects such as emission from the
Earth and Moon, the instrument's response to thermal changes, and the
instrument's response to the Earth's magnetic field. The largest detected
effects do not contribute significantly to the DMR maps; they are either on
time scales long compared to the spacecraft spin sampling (e.g. thermal gain
drifts) or have time dependence inconsistent with emission fixed on the
celestial sphere (e.g. magnetic effects). Detected and potential systematic
effects were quantitatively analyzed in detail \cite{Kogut96a}.
Data with the worst systematic contamination
(lunar emission, terrestrial emission, and thermal gain changes)
were not used in the map-making process
and constitute less than 10\% of the data in the 53 and 90 GHz channels.
The remaining data were corrected using models of each effect.
The data editing and correction parameters were conservatively chosen
so that systematic artifacts, after correction, are less than 6 $\mu$K rms (95\%
confidence upper limit) in the final DMR map in the worst channel.  
This is significantly less than the levels of the noise and celestial signals.

A dipole $T_d = 3.356$~mK anisotropy signal
(thermodynamic temperature in Galactic coordinates 
Cartesian components [X,Y,Z] = [-0.2173, -2.2451, +2.4853] mK)
is subtracted 
from the time-ordered differential data prior to forming the 4-year sky maps 
to reduce spatial gradients within a single pixel.  
 A small residual dipole remains in the maps from a combination of CMB and 
Galactic emission.
The mean signal-to-noise ratios in the 10$^\circ$ smoothed maps are approximately 
0.5, 1.5, and 1.0 for 31, 53, and 90 GHz, respectively.  
For a multi-frequency co-added map the signal-to-noise ratio is $\sim$ 2.  
This signal-to-noise level is adequate to portray an accurate overall visual 
impression of the anisotropy.  
Visual comparison of the full sky maps at each frequency, 
after averaging the A and B channels, removing the CMB dipole, 
and smoothing to 10$^\circ$ effective resolution
show coincident features.
Well off the Galactic plane these are clearly true CMB anisotropy features. 
Simulated data in combination 
with the noise appropriate to 1-, 2-, and 4-years of DMR 53 GHz observations
show the convergence of the DMR maps with the input simulated data.
Increasing years of data result in the emergence of the input large
scale features. We can be confident that the large scale features
in the 4-year DMR maps are real features rather than confusing noise.



Given the sensitivity of the 4-year DMR maps we have extended the cut
made in our previous analyses to exclude additional Galactic emission.
We use the {\it COBE} DIRBE 140 $\mu$m map as a guide to cut additional 
Galactic emission features.  The full sky DMR maps contain 6144 pixels.
An optimum Galactic cut maximizes the number of remaining
pixels while minimizing the Galactic contamination.  
This cut leaves 3881 pixels (in Galactic pixelization) 
while eliminating the strongest $\vert
b\vert>20^\circ$ Galactic emission. Moderate changes to this cut will
cause derived CMB parameters to change somewhat, but this is consistent with
the data sampling differences of real CMB anisotropy features and not
necessarily Galactic
contamination.  Likewise, derived CMB parameters also vary by the expected
amount when the maps are made in ecliptic rather than Galactic coordinates
since about 1/2 of the noise is re-binned.

%

Kogut et al. \cite{Kogut96b} examine the Galactic contamination of the high
Galactic latitude regions of the DMR maps which remain 
after the Galactic emission cut (described above). 
No significant cross-correlation is found between the DMR
maps and either the 408 MHz synchrotron map or the synchrotron map derived from
a magnetic field model \cite{Bennett92}. This places an
upper limit $T_{\rm synch} < 11 ~\mu$K (95\% confidence) on synchrotron
emission at 31 GHz.

A significant correlation is found between the DMR maps and the dust-dominated
DIRBE 140 $\mu$m map, with frequency dependence consistent with a superposition
of dust and free-free emission.  The correlation is really with
a component of Galatic emission with a spectral index of about -2.1
which could be very flat spectrum synchrotron as expected
where cosmic ray electron acceleration is actually occuring. 
We use the term free-free to stand for this component which corresponds to
a 7$^\circ$ rms free-free emission component of $7.1\pm 1.7$ $\mu$K at 53 GHz
and a dust component of $2.7\pm 1.3$ $\mu$K at 53 GHz.  Since this emission
is uncorrelated with CMB anisotropies it constitutes $<10$\% of the CMB power.
The amplitude of the correlated free-free component at 53 GHz
agrees with a noisier estimate of free-free emission
derived from a linear combination of DMR data which includes
{\it all} emission with free-free spectral dependence.
The combined dust and free-free emission contribute $10 \pm 4 ~\mu$K {\it rms}
at both 53 and 90 GHz, well below the 30 $\mu$K cosmic signal.
These Galactic signal analyses are consistent with the fact that the fitted
cosmological parameters are nearly unaffected by removal of
modeled Galactic signals \cite{Gorski96}, \cite{Hinshaw96a}
with the notable exception of the quadrupole,
which has significant Galactic contamination \cite{Kogut96b}.  A search
by Banday et al. \cite{Banday96a} finds no evidence for significant extragalactic
contamination of the DMR maps.

\subsection{Four-Year DMR Results}

{\bf Monopole $\bf \ell=0$:}  Despite the fact that the DMR is a differential 
instrument, the known motion of the {\it COBE} spacecraft about the Earth and 
the motion of the Earth about the Solar System barycenter provides a means to 
determine the CMB monopole temperature from the DMR data.  
The CMB at millimeter wavelengths is well described by a blackbody spectrum 
\cite{Mather94},\cite{Fixsen96}.  
The Doppler effect from the combined spacecraft and Earth 
orbital motions creates a dipole signal 
$T(\theta) = T_0 [ 1 + \beta \cos(\theta) + O(\beta^{2}) ]$,
where $\beta = v/c$ and $\theta$ is the angle relative to the time-dependent 
velocity vector.  The satellite and Earth orbital motions are well known 
and change in a regular fashion, allowing their Doppler signal to be separated 
from fixed celestial signals.  We fit the time-ordered data to the Doppler 
dipole and recover a value for the CMB monopole temperature,
$T_0 = 2.725 \pm 0.020$ K \cite{Kogut96a}.

{\bf Dipole $\bf \ell=1$:}
The CMB anisotropy is dominated by a dipole term attributed to the 
motion of the Solar System with respect to the CMB rest frame.
A precise determination of the dipole must account for Galactic emission 
and the aliasing of power from higher multipole orders once pixels near the 
Galactic plane are discarded.  
One can account for Galactic emission by using a linear combination of 
the DMR maps or by cross-correlating the DMR maps with template sky maps 
dominated by Galactic emission \cite{Kogut96b}.  
The high-latitude portion of the sky is fitted for a dipole with a CMB frequency 
spectrum using a pixel-based likelihood analysis \cite{Hinshaw96a}.  
Accounting for the smoothing by the DMR beam and map pixelization, 
the CMB dipole has amplitude $3.353 \pm 0.024$ mK toward Galactic coordinates 
$(l,b) = (264^\circ.26 \pm 0^\circ.33, 48^\circ.22 \pm 0^\circ.13)$,
or equatorial coordinates
$(\alpha, \delta) = (11^{\rm h} 12^{\rm m}2 \pm 0^{\rm m}8, -7^\circ.06 \pm 0^\circ.16)$
epoch J2000.

A second analysis approach utilizes a phenomenological estimate of the Galactic
foreground by examining the dipole fitted parameters as a function of
cuts in Galactic latitude. It was found that the largest source of error
in the dipole direction was reduced by using lower Galactic latitude cuts.
Using the four year data set from all six channels of the COBE Differential
Microwave Radiometers (DMR), the best-fit dipole amplitude
$3.358 \pm 0.001 \pm 0.023$ mK amplitude
in the direction
($\ell, b)=(264^\circ.31 \pm 0^\circ.04 \pm 0^\circ.16, \:+48^\circ.05 \pm 0^\circ.02 \pm 0^\circ.09)$
where the first uncertainties are statistical and the second are estimates
of the combined systematics \cite{Lineweaver96}.

These dipole measurements are consistent with previous DMR and FIRAS results.

{\bf Quadrupole $\bf \ell=2$:}
On the largest angular scales (e.g., quadrupole), Galactic emission is 
comparable in amplitude to the anisotropy in the CMB.  
The quadrupole amplitude is found by a likelihood analysis 
which simultaneously fits 
the high-latitude portion of the DMR maps for Galactic emission traced 
by synchrotron- and dust-dominated surveys and a quadrupole anisotropy 
with a thermodynamic frequency spectrum \cite{Kogut96b},\cite{Hinshaw96a}.
After correcting for the positive bias from instrument noise and aliasing,
the CMB quadrupole amplitude observed at high latitude is $Q_{rms} = 10.7\pm
3.6\pm 7.1$ $\mu$K,
where the quoted errors reflect the 68\% confidence uncertainties from random 
statistical errors and Galactic modeling errors, respectively.  The observed 
quadrupole amplitude, $Q_{rms}$, has a lower value than the quadrupole expected
from a fit to the entire power spectrum, $Q_{rms-PS}$, but whether this is a 
chance result of cosmic variance or reflects the cosmology of the Universe 
cannot be determined from {\it COBE} data.  
The 68\% confidence interval for the quadrupole amplitude, 6 $\mu$K $\le$ 
$Q_{rms}$ $\le$ 17 $\mu$K, is consistent with
the quadrupole normalization of the full power spectrum power-law fit
(discussed below): $Q_{rms-PS} = 15.3^{+3.8}_{-2.8}$ $\mu$K.

{\bf Power spectrum $\bf \ell \ge 2$:}
The simplest probe of the angular power spectrum of the anisotropy is its 
Legendre transform, the 2-point correlation function.  
The 2-point correlation function of the 4-year maps is analyzed 
by Hinshaw et al. \cite{Hinshaw96b}, where it is shown that the 2-point data are 
consistent from channel to channel and frequency to frequency.  
The data are robust with respect to the angular power spectrum.  
A Monte Carlo-based Gaussian likelihood analysis determines the most-likely 
quadrupole normalization for a scale-invariant ($n=1$) power-law spectrum.  
The results are summarized in Table \ref{Qn_fit} which also includes 
the results of 3 additional, independent power spectrum analyses, 
discussed below.  
The normalization inferred from the 2-point function is now in better agreement 
with other determinations than was the case with the 2-year data.  
The change is due to data selection: with the 2-year data, 
we only analyzed the 53 $\times$ 90 GHz cross-correlation function; 
with the 4-year data we have analyzed many more data combinations, 
including the auto-correlation of a co-added, multi-frequency map.  
This latter combination is more comparable to the data analyzed by other 
methods, and the 2-point analysis yields consistent results in that case.  
The combined 31, 53 and 90 GHz CMB rms is 29$\pm$1 $\mu$K in the
10$^\circ$ smoothed map \cite{Banday96b}, consistent with the level 
determined by the 2-point results.

It is also possible to analyze the power spectrum directly in terms of 
spherical harmonics.  However, there is considerable subtlety in this 
because the removal of the Galactic plane renders the harmonics 
non-orthonormal, producing strong correlations among the fitted amplitudes.
Wright et al. \cite{Wright96} has solved for an angular power spectrum by modifying and 
applying the technique described by Peebles \cite{Peebles73} and 
Hauser \& Peebles \cite{HauserPeebles73}
for data on the cut sphere.  They compute a Gaussian likelihood on these data 
and calibrate their results with Monte Carlo simulations.  
G\'{o}rski et al. \cite{Gorski96} explicitly construct orthonormal functions on the cut 
sphere and decompose the anisotropy data with respect to these modes.  
They form and evaluate an exact 
Gaussian likelihood directly in terms of this mode decomposition.  
The results of these analyses are summarized in Table \ref{Qn_fit}.  
Further details, including results from other data combinations are given 
in the respective papers.

Hinshaw et al. \cite{Hinshaw96a} evaluate a Gaussian likelihood directly in terms of a 
full pixel-pixel covariance matrix, a technique applied to the 2-year data 
by Tegmark \& Bunn \cite{TegmarkBunn95}.  
The results of the power-law spectrum fits are 
summarized in Table \ref{Qn_fit}.  
Hinshaw et al. \cite{Hinshaw96a} also analyze the quadrupole anisotropy 
separately from the higher-order modes, to complement 
the analysis of Kogut et al. \cite{Kogut96b}.  
They compute a likelihood for the observed quadrupole 
$Q_{rms}$, nearly independent of higher-order power, and show that it peaks 
between 6 and 10 $\mu$K, depending on Galactic model, but that its 
distribution is so wide that it is easily consistent with the 
$Q_{rms-Ps} = 15.3^{+3.8}_{-2.8} \mu$K, the value derived 
using the full power spectrum.

An important lesson from fitting with different cuts and configurations
and from our Monte Carlo simulations is that the best fitted parameters
depend both upon the random statistics of CMB fluctuations and on the
choice of cuts and fitting parameters. Table 1 in Gorski \etal \cite{Gorski96}
is indicative of the range of results obtainable using a robust
and stable approach. 

{\bf Tests for Gaussian Statistics:}
It is important to determine whether the primordial fluctuations are Gaussian.  
The probability distribution of temperature residuals should be close to 
Gaussian if the sky variance is Gaussian and the receiver noise is Gaussian.  
The receiver noise varies somewhat from pixel to pixel because the observation 
times are not all the same, but when this is taken into account the data 
appear Gaussian \cite{Smoot94}.  
There is no evidence for an excess of large deviations, 
as would be expected if there were an unknown population of point sources.   
A search for point sources in the 2-year maps found none \cite{Kogut94}.
Given the large beam of the instrument and the variance of both cosmic signals 
and receiver noise, it is still possible for interesting signals to be hidden 
in the data.  

Kogut et al. \cite{Kogut96c} compare the 4-year DMR maps to Monte Carlo simulations of 
Gaussian power-law CMB anisotropy.  The 3-point correlation function, 
the 2-point correlation of temperature extrema, and the topological genus are 
all in excellent agreement with the hypothesis that the CMB anisotropy on 
angular scales of 7$^\circ$ or larger represents a random-phase Gaussian field.
A likelihood comparison of the DMR maps against non-Gaussian $\chi^2_N$ toy 
models tests the alternate hypothesis that the CMB is a random realization
of a field whose spherical harmonic coefficients $a_{\ell m}$ are drawn from 
a $\chi^2$ distribution with $N$ degrees of freedom.  Not only do Gaussian 
power-law models provide an adequate description of the large-scale CMB 
anisotropy, but non-Gaussian models with $1 \lt N \lt 60$ are five times less 
likely to describe the true statistical distribution than the exact Gaussian 
model.

\subsection{Summary of 4-year {\it COBE} DMR CMB Measurements}
\indent
(1) The full 4-year set of {\it COBE} DMR observations is analyzed and full sky
maps have been produced \cite{Bennett96}. 
The typical signal-to-noise ratio in a 10$^\circ$ smoothed frequency-averaged 
map is $\sim 2$, enough to provide a visual impression of the anisotropy.

(2) The DMR (despite its being a differential instrument) finds a
CMB monopole temperature of $T_0 = 2.725 \pm 0.020$ K \cite{Kogut96a}.  
This is in excellent agreement
with the {\it COBE} FIRAS precision measurement of the spectrum of the CMB,
$T_0 = 2.728 \pm 0.002$ K \cite{Fixsen96}. 

(3) The CMB dipole from DMR has amplitude $3.358 \pm 0.024$ mK toward Galactic 
coordinates 
$(l,b) = (264^\circ.31 \pm 0^\circ.16, 48^\circ.05 \pm 0^\circ.10)$, or 
equatorial coordinates
$(\alpha, \delta) = (11^{\rm h} 12^{\rm m}2 \pm 0^{\rm m}8, -7^\circ.06 \pm 0^\circ.16)$
epoch J2000.  This is consistent with the dipole amplitude and direction
derived by {\it COBE} FIRAS \cite{Fixsen96}.

(4) The 95\% confidence interval for the observed $\ell=2$ quadrupole amplitude is 
4 $\mu$K $\le$ $Q_{rms}$ $\le$ 28 $\mu$K.  This is consistent with the value
predicted by a power-law fit to the power spectrum yields a quadrupole 
normalization of: 
$Q_{rms-PS} = 15.3^{+3.8}_{-2.8}$ $\mu$K \cite{Kogut96b}; \cite{Hinshaw96a}.

(5) The power spectrum of large angular scale CMB measurements is consistent
with an $n=1$ power-law \cite{Gorski96}, \cite{Hinshaw96a}, \cite{Wright96}.  
If the effects of a standard cold dark matter model are
included, {\it COBE} DMR should find $n_{eff}\approx 1.1$ for a $n=1$ 
universe. 
With full use of the multi-frequency 4-year DMR data, 
including our estimate of the effects of Galactic emission, 
we find a power-law spectral index of $n=1.2\pm 0.3$ and a 
quadrupole normalization $Q_{rms-PS}=15.3^{+3.8}_{-2.8}$ $\mu$K. 
For $n=1$ the best-fit normalization is 
$Q_{rms-PS}\vert_{n=1}=18\pm 1.6$ $\mu$K.  Differences in the derived values of
$Q$ and $n$ between various analyses of DMR data are much more dependent on the
detailed data selection effects than on the analysis technique.

(6) The DMR anisotropy data are consistent with Gaussian statistics.
Statistical tests prefer Gaussian over other toy statistical models
by a factor of $\sim 5$ \cite{Kogut96c}.

\begin{table}
\begin{minipage}{8cm}
\begin{tabular}{lccc}
\hline
Technique & $n~^a$ & $Q_{rms-PS}~^b$ & $Q_{rms-PS|n=1}~^c$ \\
        & & ($\mu$K) & ($\mu$K) \\
\hline
\multicolumn{4}{c}{No Galaxy Correction$~^d$} \\
2-point correlation function~\cite{Hinshaw96b} & ---     & ---     & $17.5^{+1.4}_{-1.4}$ \\
Orthogonal functions~\cite{Gorski96} & $1.21^{+0.24}_{-0.28}$ &
                     $15.2^{+3.7}_{-2.6}$ & $17.7^{+1.3}_{-1.2}$ \\
Pixel temperatures~\cite{Hinshaw96a} & $1.23^{+0.26}_{-0.27}$ &
                     $15.2^{+3.6}_{-2.8}$ & $17.8^{+1.3}_{-1.3}$ \\
Hauser-Peebles cut sky~\cite{Wright96} & ---  & --- & ---  \\
\hline
\multicolumn{4}{c}{Internal Combination Galaxy Correction$~^e$ } \\
2-point correlation function~\cite{Hinshaw96b} & ---     & ---     & $16.7^{+2.0}_{-2.0}$ \\
Orthogonal functions~\cite{Gorski96} & $1.11^{+0.38}_{-0.42}$ &
                     $16.3^{+5.2}_{-3.7}$ & $17.4^{+1.8}_{-1.7}$ \\
Pixel temperatures~\cite{Hinshaw96a} & $1.00^{+0.40}_{-0.43}$ &
                     $17.2^{+5.6}_{-4.0}$ & $17.2^{+1.9}_{-1.7}$ \\
Hauser-Peebles cut sky~\cite{Wright96} & $1.62^{+0.44}_{-0.50}$ &
                     ---                 & $19.6^{+2.5}_{-2.5}$\\
\hline
\end{tabular}
\begin{tabular}{l}
$^a${Mode \& 68\% confidence range of the projection of 
the 2-d likelihood, $L(Q,n)$, on $n$} \\

$^b${Mode \& 68\% confidence range of the projection of 
the 2-d likelihood, $L(Q,n)$, on $Q$} \\

$^c${Mode \& 68\% confidence range of the slice of the 
2-d likelihood, $L(Q,n)$, at $n=1$} \\

$^d${Formed from the weighted average of all 6 channels} \\



$^e${from a linear combination of 6 channel maps
canceling free-free emission \cite{Kogut96b}} \\
\end{tabular}
\end{minipage}
\caption{Summary of DMR 4-Year Power Spectrum Fitting Results.}
\label{Qn_fit}
\end{table}

\subsection{COBE Conclusions}

The COBE-discovered \cite{Smoot92}
higher-order ($\ell \ge 2$) anisotropy is interpreted
as being the result
of perturbations in the energy density of the early Universe,
manifesting themselves at the epoch of the CMB's last scattering.
These pertubations are the seeds of large scale structure formation
and are relics from processes occurring in the very early Universe
at extremely high energies.
In the standard scenario the last scattering of cosmic background photons 
takes place at a redshift of approximately 1100, at which epoch the large
number of photons was no longer able to keep the hydrogen sufficiently ionized.
The optical thickness of the cosmic photosphere is roughly $\Delta z \sim 100$
or about 10 arcminutes, so that features smaller than this size are damped.
Observations of the CMB anisotropy power spectrum can reveal to us
much of the interesting history of the early Universe 
and so a great deal of effort has gone into its observation.

\section{Current Anisotropy Power Spectrum}
On the order of ten experiments have now observed CMB anisotropies.
Anisotropies are observed on angular scales larger than the
minimum 10$^\prime$ damping scale (see Figure 3)
and are consistent with those expected from an initially scale-invariant power
spectrum 
of potential and thus metric fluctuations.
It is believed that the large scale structure in the Universe
developed through the process of gravitational instability
where small primordial perturbations in energy density were amplified
by gravity over the course of time.
The initial spectrum of density perturbations
can evolve significantly in the epoch $z > 1100$ for causally connected
regions (angles $\lsim 1^\circ\;\Omega_{tot}^{1/2}$).
The primary mode of evolution
is through adiabatic (acoustic) oscillations, leading to a series
of peaks that encode information about
the perturbations and geometry of the universe, as well as
information on $\Omega_0$, $\Omega_B$, $\Omega_\Lambda$ (cosmological
constant),
and $H_0$ \cite{scott95}.
The location of the first acoustic peak
is predicted to be at $\ell \sim 220 \;\Omega_{tot}^{-1/2}$ or
$\theta \sim 1^\circ \;\Omega_{tot}^{1/2}$
and its amplitude increases with increasing $\Omega_B$.

\begin{figure}
\centerline{\epsfxsize= 9 cm \epsfbox[156 235 459 456]{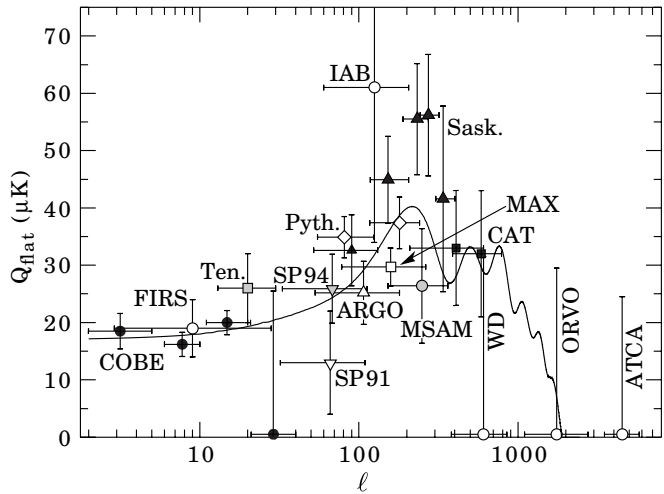}}
  \caption
{Current status of CMB anisotropy observations: 
Plotted are the quadrupole amplitudes
for a flat (unprocessed scale-invariant spectrum of primordial perturbations,
i.e., a horizontal line) anisotropy spectrum
that would give the observed results for the experiment.
The vertical error bars represent estimates of 68\% CL, while
the upper limits are at 95\% CL.  Horizontal bars
indicate the
range of $\ell$ values sampled.
The curve indicates the expected spectrum for a standard CDM model
($\Omega_0 = 1$, $\Omega_B = 0.05$, $h = 0.5$), 
although true comparison with models should involve convolution 
of this curve with each experimental filter function.  
}
\label{scott}
\end{figure}

\begin{figure}
\centerline{\epsfxsize= 9 cm \epsfbox[18 144 592 718]{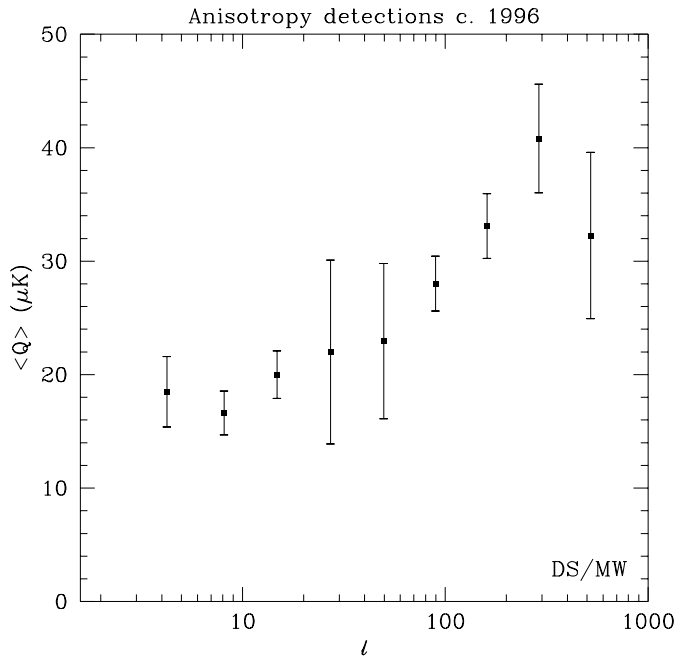}}
  \caption
{Current status of CMB anisotropy observations: 
Due to the overlapping sets of data Douglas Scott
and Martin White have formed band-averages over experiments.
Plotted are the quadrupole amplitudes
for a scale-invariant spectrum 
that would give the experiment band averages.
The vertical error bars represent estimates of 68\% CL.
}
\label{scott2}
\end{figure}


Figure 4 shows the theoretically predicted power spectrum for a sample of models,
plotted as $\ell(\ell+1) C_\ell$ versus $\ell$ which is the power per
logarithmic interval in $\ell$ or, equivalently, the two-dimensional
power spectrum.
If the initial power spectrum of perturbations is the result of
quantum mechanical fluctuations produced and
amplified during inflation, then
the anisotropy spectrum and the fractional contribution from
density (scalar) and gravity wave (tensor) perturbations
are coupled.
If the energy scale of inflation at the appropriate epoch is at
the level of $\simeq 10^{16}$GeV, then detection of gravitons is possible,
as well as partial reconstruction of the inflaton potential.
If the energy scale is $\lsim 10^{14}$GeV, then density
fluctuations dominate and less constraint is possible.
(See CMB theory lectures for more background.)

\begin{figure}
\centerline{\epsfxsize= 9 cm \epsfbox[142 161 432 383]{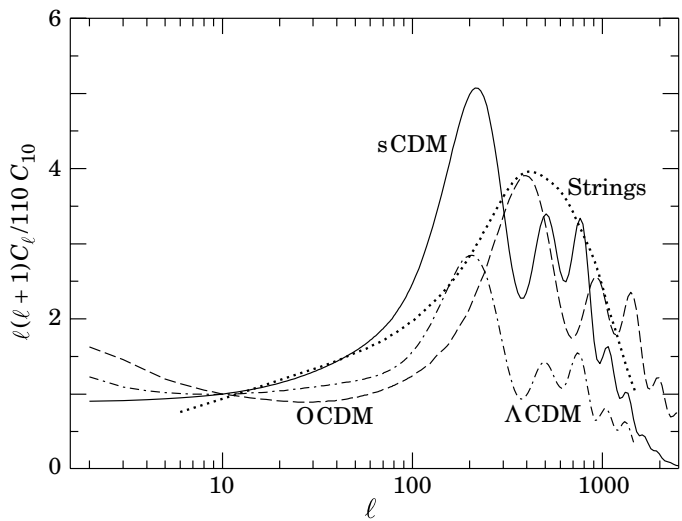}}
\caption{Examples of theoretically predicted $\ell (\ell +1) C_\ell$
or CMB anisotropy power spectra.
{\bf sCDM} is the standard cold dark matter model 
with $ h = 0.5$ and $\Omega_B = 0.05$.
{\bf $\Lambda$CDM} is a model 
with $\Omega_{tot} = \Omega_\Lambda +\Omega_0 = 1$
where $\Omega_\Lambda=0.3$ and $h=0.8$.
{\bf OCDM} is an open model with $\Omega_0 = 0.3$ and $h = 0.75$.
(See White (1996) 
for models).
{\bf Strings} is a model where cosmic strings are the primary source
of large scale structure [1] 
The plot indicates that precise measurements
of the CMB anisotropy power spectrum could distinguish between current
models.}
\label{smoot_theory}
\end{figure}


Fits to data over smaller angular scales are often quoted as the expected
value of the quadrupole $\left\langle Q\right\rangle$
for some specific theory, e.g. a model with power-law
initial conditions (primordial
density perturbation power spectrum $P(k)\propto k^n$).
The full 4-year COBE DMR data give
$\left\langle Q\right\rangle=15.3^{+3.7}_{-2.8}\;\mu$K, after projecting out
the slope dependence, while the best-fit slope is
$n=1.2\pm0.3$, and for an $n=1$
(scale-invariant potential perturbation) spectrum
$\left\langle Q\right\rangle
(n=1)=18\pm1.6\,\mu$K \cite{Bennett96}, \cite{Gorski96}.
 The conventional notation
is such that $Q_{\rm rms}^2/T_\gamma^2=5C_2/4\pi$.

Only somewhat weak conclusions can be drawn based on the current
smaller angular scale
data (see Figure 3).  However, new data are being acquired at an
increasing rate.  With future experiments and the prospect of next generation
satellite missions: MAP and Planck, 
a precise measurement of the CMB anisotropy power spectrum
is possible and likely, allowing us to decode the information that it
contains \cite{Knox95},\cite{Jungman96}.

\section{Current and Near-Term Experiments}

Many groups have been and are working to measure the anisotropy. 
Though some are focussed on large angular scales at frequencies not observed 
with DMR, most concentrate on smaller angular scales. 
Table 2 contains a list for recent, current and planned experiments
compiled by L. Page \cite{Page97}.

\begin{table}[p]
\begin{center}
\caption{Recently Completed, Current and Planned Anisotropy
Experiments}
\begin{tabular}{lccccl}
\hline
Experiment & Resolution & Frequency & Detectors & Type & Groups\\
\hline
ACE(c)\cite{ACE}          & $0.2^{\circ}$ & 25-100 GHz & HEMT & C/B  & UCSB \\
APACHE(c)\cite{APACHE}  & $0.33^{\circ}$ & 90-400 GHz & Bol & C/G  & Bologna, Bartol \\
                                                 &&&&& Rome III \\
ARGO(f)\cite{ARGO}  & $0.9^{\circ}$ & 140-3000 GHz & Bol & C/B  & Rome I \\
ATCA\cite{ATCA}    & $0.03^{\circ}$ & 8.7 GHz & HEMT & I/G & CSIRO \\
BAM(c)\cite{BAM}     & $0.75^{\circ}$ & 90-300 GHz & Bol & C/B & UBC, CfA \\ 
Bartol(c)\cite{TR2} & $2.4^{\circ}$  & 90-270 GHz  & Bol & C/G & Bartol \\ 
BEAST(p)\cite{ACE}      & $0.2^{\circ}$ & 25-100 GHz & HEMT & C/B  & UCSB \\
BOOMERanG(p)\cite{BOOMERanG}   & $0.2^{\circ}$ & 90-400 GHz & Bol & C/G 
                                                 & Rome I, Caltech, \\
                                                 &&&&&  UCB, UCSB \\
CAT(c)\cite{CAT}    & $0.17^{\circ}$ & 15 GHz & HEMT & I/G & Cambridge \\
CBI(p)\cite{CBI}    & $0.0833^{\circ}$  & 26-36 GHz & HEMT & I/G & Caltech, Penn. \\
FIRS(f)\cite{FIRS}  & $3.8^{\circ}$ & 170-680 GHz & Bol & C/B & Chicago, MIT,\\ 
                                                  &&&&&Princeton,\\
                                                  &&&&&NASA/GSFC \\
HACME/SP(f)\cite{HACME}    & $0.6^{\circ}$  & 30 GHz & HEMT & C/G & UCSB \\
IAB(f)\cite{IAB}    & $0.83^{\circ}$ & 150 GHz & Bol & C/G & Bartol \\
MAT(p)\cite{MAT}    & $0.2^{\circ}$  & 30-150 GHz & HEMT/SIS &
C/G & Penn, Princeton \\
MAX(f)\cite{MAX}    & $0.5^{\circ}$  & 90-420 GHz & Bol & C/B & UCB, UCSB \\
MAXIMA(p)\cite{MAXIMA}   & $0.2^{\circ}$  & 90-420 GHz & Bol & C/B & UCB, Rome I\\
&&&&& Caltech, UCSB \\
MSAM(c)\cite{MSAM}    & $0.4^{\circ}$  & 40-680 GHz & Bol & C/B & Chicago, Brown, \\
                                                &&&&& Princeton, \\
                                                &&&&& NASA/GSFC \\
OVRO 40/5(c)\cite{OVRO} & $0.033^{\circ},0.12^{\circ}$ & 15-35 GHz & HEMT & C/G & Caltech, Penn \\ 
PYTHON(c)\cite{PYTHON} & $0.75^{\circ}$ & 35-90 GHz & Bol/HEMT & C/G & Carnegie Mellon\\
						&&&&& Chicago, UCSB\\ 
QMAP(f)\cite{QMAP} & $0.2^{\circ}$ & 20-150 GHz  & HEMT/SIS & C/B & Princeton, Penn \\ 
SASK(f)\cite{SASK} & $0.5^{\circ}$ & 20-45 GHz  & HEMT & C/G & Princeton \\ 
SuZIE(c)\cite{SuZie} & $0.017^{\circ}$ & 150-300 GHz  & Bol & C/G & Caltech \\ 
TopHat(p)\cite{TopHat} & $0.33^{\circ}$ & 150-700 GHz & Bol & C/B & Bartol, Brown,\\ 
                                       &&&&& DSRI,Chicago, \\
                                       &&&&& NASA/GSFC\\
Tenerife(c)\cite{Tenerife}  & $6.0^{\circ}$  & 10-33 GHz  & HEMT & C/G & 
NRAL, Cambridge \\ 
VCA(p)\cite{VCA} & $0.33^{\circ}$ & 30 GHz  & HEMT & I/G & Chicago \\ 
VLA(c)\cite{VLA} & $0.0028^{\circ}$ & 8.4 GHz  & HEMT & I/G & Haverford, NRAO \\ 
VSA(p)\cite{VSA} & -- & 30 GHz  & HEMT & I/G & Cambridge \\ 
White Dish(f)\cite{WD} & $0.2^{\circ}$ & 90 GHz  & Bol & C/G & Carnegie Mellon \\ 
\hline
\end{tabular}
\end{center}
\begin{enumerate}
\item For ``Type'' the first letter distinguishes between configuration
or interferometer, the second between ground or balloon.
\item An ``f'' after the experiment's name means it's finished;
a ``c'' denotes current; a ``p'' denotes planned, building may 
be in progress but there is no data yet.
\end{enumerate}
\end{table}

Rather than review all the experiments, we focus here on 
a representative sample.

\subsection{MAX/MAXIMA/BOOMERANG}
The MAX/MAXIMA/BOOMERANG payloads are representative of current
and currently planned balloon-borne missions.

\subsubsection{MAX}
The Millimeter-wave Anisotropy eXperiment (MAX) is a balloon-borne
bolometric instrument which observes at multiple frequencies
with high sensitivity on the 0.5$^\circ$ angular scale.
MAX has completed five flights detecting significant CMB fluctuations
\cite{Fischer92}, \cite{Alsop92}, \cite{Meinhold93}, 
\cite{Devlin94}, \cite{Clapp94}, \cite{Tanaka96}, \cite{Lim96}.

The MAX instrument consists of an off-axis Gregorian telescope
and a bolometric photometer mounted on an attitude-controlled balloon-borne
platform which makes measurements at an altitude of 36 km.
The Gregorian telescope consists of a 1-meter primary
and a nutating elliptical secondary.
The underfilled optics provides a 0.55$^\circ$ FWHM beam when focused
and aligned.
The 5.7 Hz nutation of the secondary modulates the beam on the sky
sinusoidally though $\pm 0.68^\circ$ and the attitude control
sweeps the beam over a $6^\circ$ or $8^\circ$ path
and back in about 108 seconds,
producing about 15 to 20 independent temperature differences on the sky.
Depending upon the time of observation and location of the region under observation
sky rotation can cause the observed region to be in the shape of a bow-tie.

On flights 4 \&\ 5 the single-pixel four-band bolometric receiver
featured negligible sensitivity to radio frequency interference and
an adiabatic demagnetization refrigerator to cool the photometer to 85 mK.
The dichroic photometer used for MAX has 
($\delta \nu / \nu$) of 0.57, 0.45, 0.35, and 0.25 filter bands
at 3.5, 6, 9, and 15 cm$^{-1}$.
MAX covers the high frequency side
of the window formed by galactic dust emission rising at higher
frequencies and Galactic synchrotron and free-free emission increasing
at lower frequencies.
The 15 cm$^{-1}$ channel acts as a guard against Galactic dust
and atmospheric emission.
The multiple frequencies have sufficient redundancy
to provide confidence that the signal is CMB and not a foreground
or systematic effect.

MAX is calibrated both by an on-board commandable membrane and
by observations of planets, usually Jupiter.
The two techniques agree at roughly the 10\%\ level.
The calibration is such that the quoted temperature difference
is the real temperature difference on the sky.

MAX makes deep CMB observations (typically one hour) on regions generally
selected to be low in dust contrast and total emission and free from
known radio sources. MAX has made observations on five flights. 
The data from most of the scans are in good agreement but the
scan of the $\mu$-Pegasi region is significantly lower than the rest.
A combination of all the data seems to be coming out at an intermediate value
between GUM and $\mu$-Pegasi regions and may all be consistent
with coming from a single parent population \cite{Tanaka97}.

The center of the scan is the same for the three observations of GUM
(the star Gamma Ursae Minoris) but the relative geometry is such that
the three scans made bow-tie patterns which cross at the star.
White and Bunn \cite{WhiteBunn95} have made use of this fact to construct
a two dimensional map of the region which is roughly $10^\circ \times 5^\circ$.
The title of their paper is ``A First Map of the CMB at 0.5$^\circ$ Resolution".
Since then a map covering 180 square degrees was generated by Tegmark et al.
\cite{Tegmark96} using the Saskatoon data.

Making maps is clearly the appropriate approach for the current
generation of new experiments.
MAX is evolving to new systems MAXIMA and BOOMERANG, which are designed
and constructed for the goal of getting the power spectrum
around the first ``Doppler'' peak and further and making maps covering
a significant portion of the sky.

\subsubsection{MAXIMA}
MAXIMA stands for MAX imaging system.
The current one-dimensional scans are very useful data
for the discovery phase of CMB anisotropy research.
Soon progress will depend upon the availability of two-dimensional maps
of low galactic foreground regions (low dust in this case)
with several hundred pixels
so that sampling variance is less important (see section 3).
In addition one can look for properties of the sky which are not
predicted by theories and could be overlooked in statistical analyses.
It also makes it possible to catalog features for comparison to or motivation
of other experiments.

Under the auspices of the NSF Center for Particle Astrophysics
a collaboration consisting of groups from
the University of California at Berkeley,
Caltech, the University of Rome, and the IROE-CNR Florence
have begun work and made good progress on the new systems for MAXIMA and 
BOOMERANG. The plan is to have a combination of northern hemisphere flights
of MAXIMA and BOOMERANG and a Long Duration Balloon (LDB) flight 
of BOOMERANG from Antarctica.

\begin{figure}
\centerline{\epsfxsize= 13 cm \epsfbox[ 10 12 600 779]{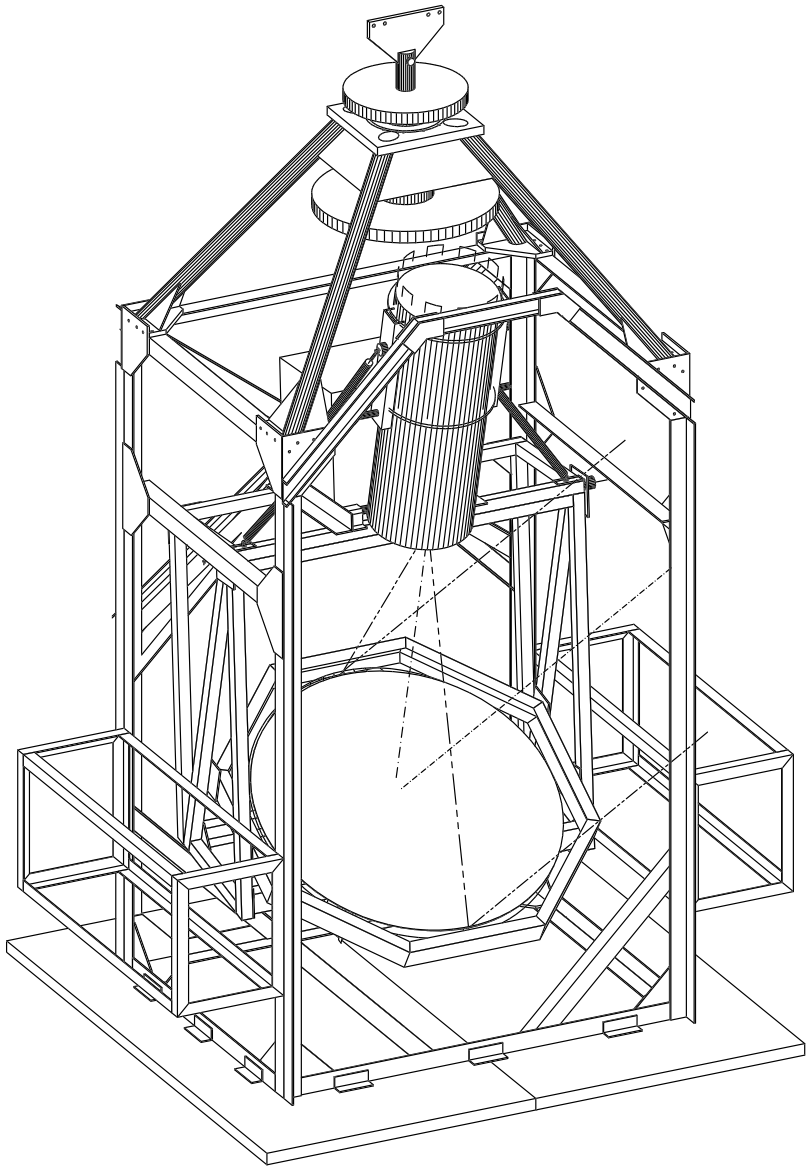}}
\caption
{A schematic drawing of the MAXIMA gondola system 
showing three sample rays from the sky reflecting
from the nodding primary, coming to the prime focus
and entering the dewar containing cold optics and bolometer detectors.
Also show are the gondola frame work with the angular momentum wheels
on top and two side boxes holding the electronics for the pointing
system and the detector signal processing.
The first stage of detector electronics is inside the dewar,
the next stage in an RF shielded backpack attached to the dewar.
}
\label{MAXgondola}
\end{figure}

To make an imager a new optical system was necessary.
The primary mirror for MAXIMA is a 1.3-meter, off-axis, light-weight primary 
mirror.
The primary will be modulated which allows a much larger beam chop angle
on the sky with less spill over and thus more pixels in the focal plane.
Cold secondary and tertiary mirrors provide a cold Lyot stop
and the field-of-view required for the array of 12 arcminute pixels.
The geometrical aberrations in the center of the field-of-view
are less than 10 arcminutes.

A larger primary mirror requires a larger gondola which is now constructed.
The chop angle can both be increased and varied allowing the instrument
to sample the shape of the power spectrum over the range
$40 < \ell < 1000$.

An additional feature is new detector electronics with AC coupling
in order to allow linear scanning in a total power mode,
making maps and power spectrum measurements directly.
This approach is different than that of making
a number of different window functions. 
The idea is to use a scan or raster scan of the CMB anisotropies
on the sky directly rather than obtaining a set of differences
at different chop angles. One is thus mapping directly and
measuring the power spectrum as the fourier transform of the data.
At this stage the instrument is designed to operate in this mode by
scanning the primary mirror in a sawtooth pattern rapidly (3 Hz) 
and more slowly moving the entire gondola in azimuth
to cover a larger angle.

\begin{figure}
\centerline{\epsfxsize= 9 cm \epsfbox[ 29 38 299 380]{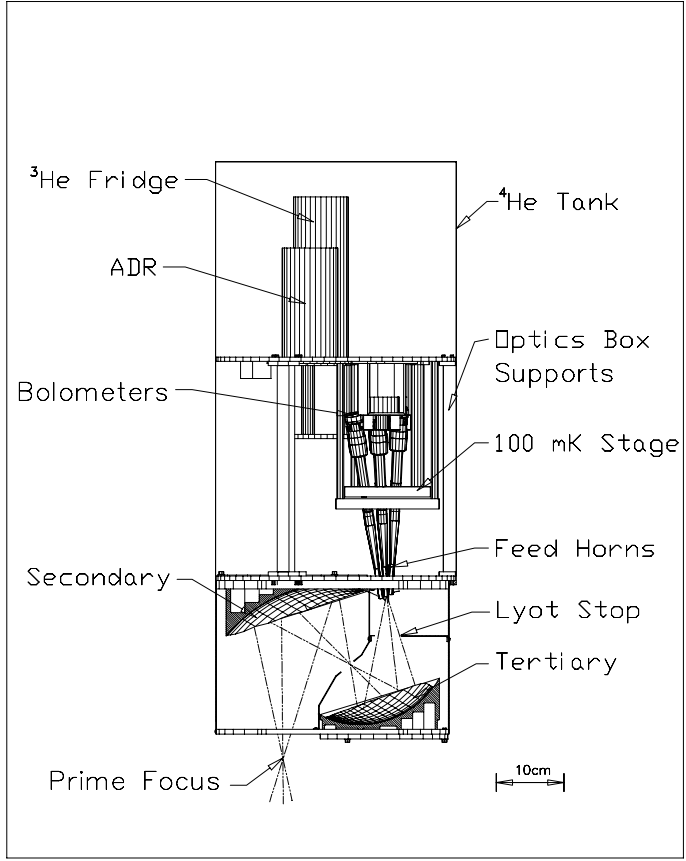}}
  \caption
{The cold optical system for MAXIMA. 
Both MAXIMA and BOOMERANG use fast off-axis LHe-cooled secondary and 
tertiary mirrors to reduce optics emissions.
The design incorporates cold black baffles and a cold Lyot stop
which controls the illumination on the primary mirror
giving smaller offsets and better control of spillover than can be achieved
in optical systems that are not re-imaged.
The figure also shows the location of the feedhorns and the bolometers
as well as the Adiabatic Demagnitization Refridgerator (ADR) and He3
fridges used to maintain the bolometers at 100 mK.
The optics are shown in cross-section for clarity.
}
\label{MAXCO}
\end{figure}

\begin{figure}
\centerline{\epsfxsize= 9 cm \epsfbox[18 144 592 718]{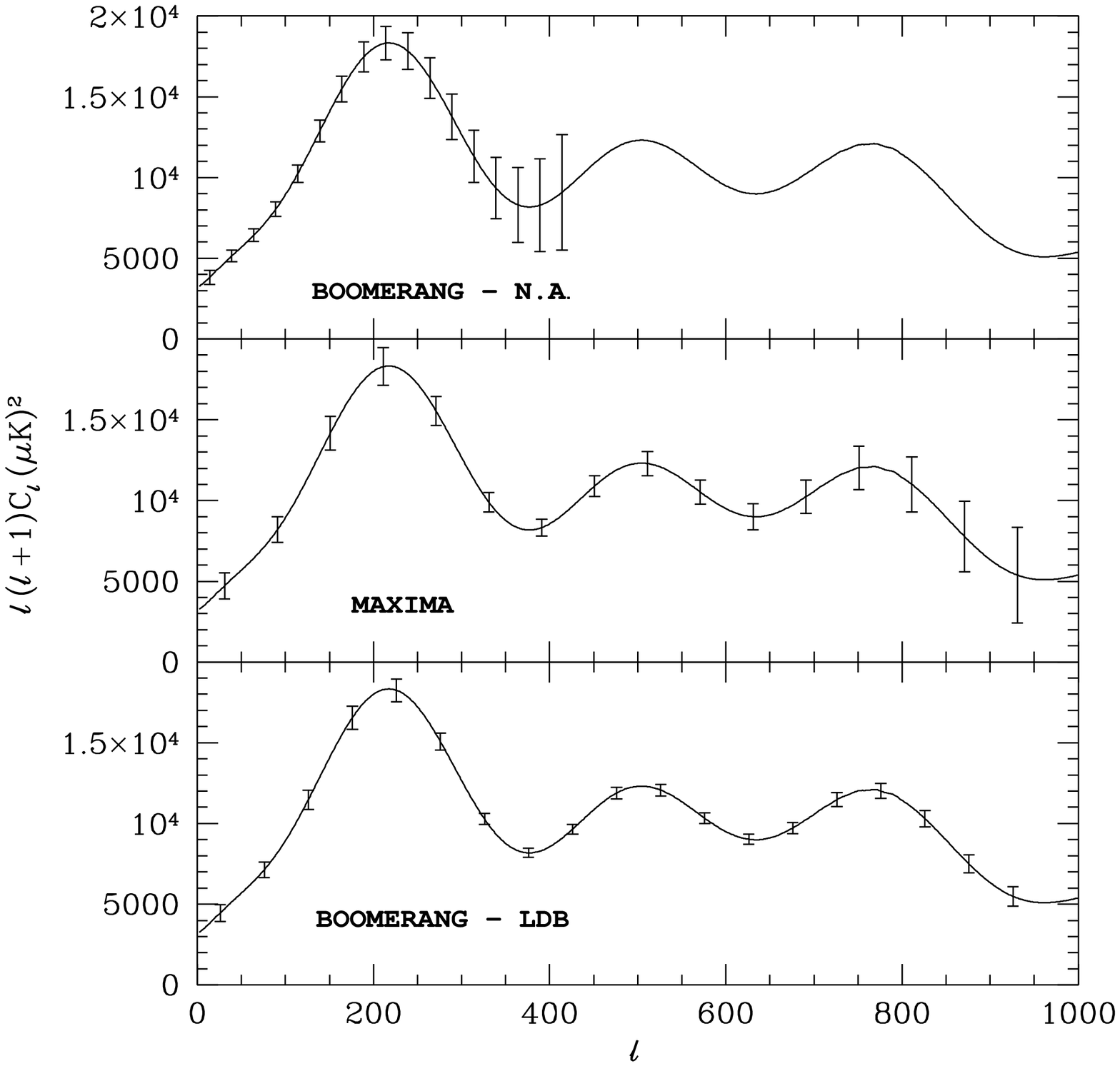}}
  \caption
{Estimates of the accuracy with which each of the three proposed flights
for MAXIMA and BOOMERANG will determine the power spectrum of CMB anisotropy.
The sold line in each frame represents the standard cold dark matter
($\Omega = 1, ~\Omega_b = 0.05, ~ h = 50$, no reionization) power spectrum.
The spacing of the error flags indicates the resolution provided by each
experiment; the amplitude of the error flags indicate one standard deviation
uncertainty and includes the effects of instrumental noise, sky coverage,
and differences strategy.
}
\label{MAXCL}
\end{figure}

Another major change will be going from a single pixel four-frequency
photometer to a fourteen-pixel receiver.
This will allow taking data at eight times the rate and thus make
two-dimensional mapping feasible. 
The receiver design has been completed and it and the new cold
optics being mounted in the new large dewar as show in Figure \ref{MAXCO}.
The bolometers have a spider-web (silicon nitride micromesh) substrate
so that cosmic ray transient occurrences will be reduced by more
than an order of magnitude.
The first flight of the new gondola was September 1995
and we anticipate a flight with the arrary receiver in the summer/fall 1997.
We can anticipate that within three years MAXIMA will have made maps and
will have measured the anisotropy power spectrum around the location
of the first doppler peak. Figure \ref{MAXCL} indicates an estimate
of the accuracy of the power spectrum determination.

\subsubsection{BOOMERANG}
BOOMERANG is equivalent to the long-duration balloon-borne version
of MAXIMA and an intermediate step toward the bolometer space mission,
the Planck HFI.
Plans call for a northern hemisphere flight in June 1997 
followed by a many-day flight circumnavigating Antarctica
in the austral summer beginning December 1998.
BOOMERANG will move more directly towards mapping a significant region
of the sky. 
The BOOMERANG focal plane contains 8 pixels: 
four multiband photometers (6, 9 and 14 cm$^{-1}$)
and four monochromatic channels (3 cm$^{-1}$). 
The diffraction limited angular resolution is 12$^\prime$
above 6 cm$^{-1}$ and 20$^\prime$ at 3 cm$^{-1}$. 
In total power mode, the largest resolution is limited only 
by the length of a scan. 

\begin{figure}
\centerline{\epsfxsize= 6 cm \epsfbox[34 56 244 309]{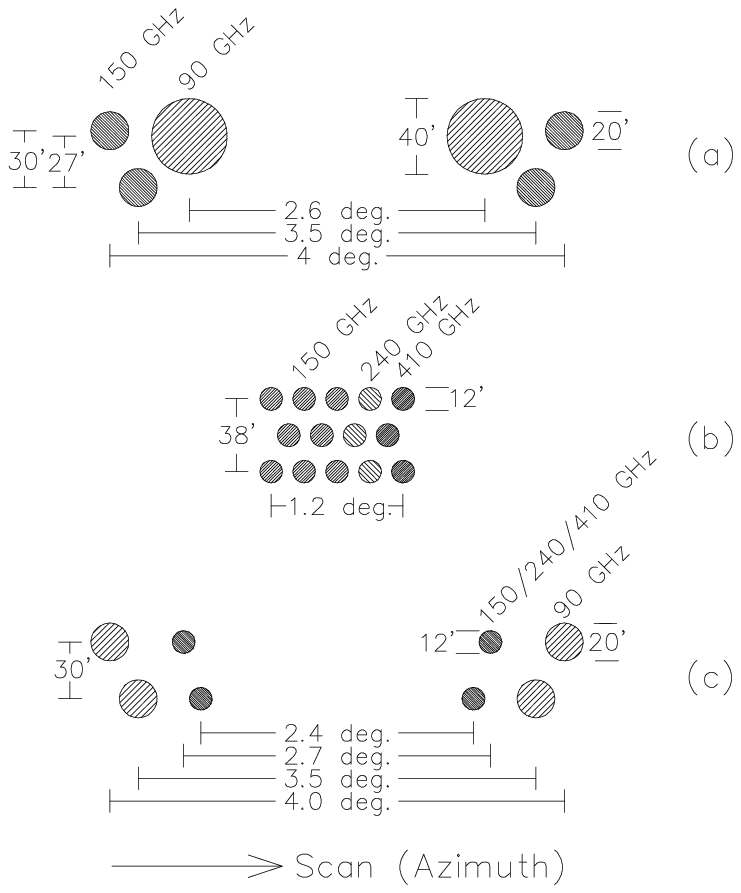}}
\caption
{Layout of the three BOOMERANG and MAXIMA focal planes.
The focal planes are:
(a) BOOMERANG North American flight,
(b) MAXIMA, and 
(c) BOOMERANG LDB (Antarctica).
Circles indicate the relative position and FWHM of the beams 
projected on the sky.
Each circle represents a dual-polarization, high efficiency feed,
with the exception 12$^\prime$ beams shown in (c), 
which represent multi-frequency photometers of the type flown previously on MAX.
Each experiment is scanned in azimuth, which is horizontal in the figure.
}
\label{BOOMF}
\end{figure}

The high cosmic ray flux over the Antarctic requires detectors which are
insensitive to cosmic rays. 
``Spider web" bolometers have been developed specifically to minimize
the effect of cosmic rays on the detector.
These bolometers are called composite because
the functions of absorbing radiation and
measuring the temperature increase are separated. 
The absorber is the ``spider web" and a thermometer is a neutron
transmutation doped (NTD 14) thermistor.
Spider web bolometers for BOOMERANG 
will have a Noise Equivalent Power 
(NEP) of $\sim~1 \times 10^{-17}$ W/Hz$^{1/2}$. 

The electrical circuit of BOOMERANG is split in subcircuits which reside at
different temperatures.
There are bolometers at 300 mK, which are AC-biased at ~200 Hz and dual JFET
source followers, providing a low impedance line going out of the cryostat. 
At 300 K, the signal is preamplified, demodulated and filtered through a
small bandwidth, thus enabling the detection of signals of the order of 
nanovolts. 

The BOOMERANG North American flight will produce a map covering 
a region near the North Celestial Pole.
It is about 10\% of the full sky.
This region will be mapped by sweeping in azimuth between 60 degrees
and 90 degrees North latitude while holding the elevation constant; 
after 12 hours, the rotation of the earth will produce a map of half 
of a circular region centered on the NCP. 
The North American flight will produce the first total power map of a
significant fraction of the northern sky. 

The region to be mapped by the Antarctic flight 
is centered on the Southern Hole, 
a region of exceptionally low galactic dust emission in the southern sky. 
During the Antarctic summer, the anti-solar direction passes through the 
Southern Hole; the BOOMERANG scans will be centered on the antipode of 
the sun's path to minimize sidelobe response and thermal response from the sun. 

The Antarctic flight will consist of three scans, each lasting five days: 
two fast scans of adjoining $30^\circ\times 60^\circ$,
which cover ~10\%\ of the sky, 
and a deep integration of a $12^\circ\times 12^\circ$ subset of this region. 
The fast scans will result in 130 000 pixels (12 arcmin) with a sensitivity 
of 20~$\mu$K/pixel (thermodynamic temperature), obtained 
by sweeping the telescope at 1 degree/s in azimuth while
varying the elevation through 40 degrees over a 24 hour period. 
The deep integration region will be mapped with a sensitivity of 
$\sim 10 ~\mu$K/pixel. 
This scan will test for systematic variations in the data and will 
serve as a diagnostic for the fast scans. 

BOOMERANG will complement the DMR with high sensitivity
measurement of CMB power on scales between 12 arcmin and $\sim 10^\circ$. 
In Figure \ref{MAXCL}, the error bars represent the $\pm 1\sigma$ limit
which BOOMERANG will determine for a standard cold dark matter model.
The BOOMERANG-MAXIMA program will return excellent scientific data and 
will be a good test of the instrumentation and techniques for Planck.

\subsection{MSAM/TOPHAT}

MSAM/TopHat is a collaboration working on a series of experiments to measure 
the medium-scale anisotropy of the cosmic microwave background radiation. 
The collaboration includes researchers at the Bartol Research Institute, 
Brown University, the University of Chicago, the Danish Space
Research Institute, and NASA/Goddard Space Flight Center. 
MSAM/TOPHAT program is similar to MAX/MAXIMA/BOOMERANG program at present. 

\subsubsection{MSAM}
MSAM is an acronym for Medium Scale Anisotropy Measurement.
A notable difference between MSAM and MAX has been that MSAM used 
a three-position chop analyzed either as a triple beam or double beam 
(two chop angles on the sky) observation. 
MSAM angular resolution is 0.5$^\circ$ between 5 and 23 cm$^{-1}$
(150 and 700 GHz or wavelengths 0.4 to 2.0 mm).
MSAM has had three flights (June 1992, May 1994, June 1995) all from
Palestine, Texas \cite{Cheng95}.

MSAM-I: The first phase of the Medium-Scale Anisotropy Measurement (MSAM-I)
probed CMBR anisotropy at 0.5 degree angular scales between 5 and 23 cm$^{-1}$ 
(150 and 700 GHz, 0.4 and 2.0 mm). The first flight of this package 
(June 1992 from Palestine, TX) has resulted in a detection of 
$0.5 \times 10^{-5}  < \delta T/T < 2 \times 10^{-5}$. 
Fluctuations at these angular scales are believed to be the precursors of the 
largest structures we observe today. 
This level of anisotropy is at the lower end of the predicted values from 
standard Cold Dark Matter theories of structure formation. 
The interpretation of the results, however, is complicated by the presence of 
two point sources in the data. 
This hints at the possibility of a previously unsuspected population of objects 
which will challenge anisotropy measurements at these sensitivity levels. 
Such sources may be distinguished by the use of multiple spectral bands 
bracketing the peak of the CMBR, such as those in MSAM-I and MSAM-II. 
One can also interpret the two extra peaks as CMB fluctuations and then
the data are in good agreement with the anisotropy predicted by SCDM.

MSAM-II 
is the second phase of the Medium-Scale Anisotropy Measurement. 
An adiabatic demagnetization refrigerator (ADR) is used to cool 
monolithic silicon bolometers to 100mK in a new radiometer.
The new radiometer has expanded frequency coverage
in 5 spectral bands between 2.3 and 5 cm$^{-1}$ (70 to 150 GHz
or wavelengths 2.0 to 4.3 mm). The instrument is expected to improve
the signal-to-noise ratio by about a factor of three over the
previous results.

MSAM observations are along a ring surrounding the north celestial pole 
with a 20$^\prime$ beam on the sky giving sensitivity to
the medium angular-scale power spectrum of the radiation.

\subsubsection{TOPHAT}
TOPHAT is conceived as a long-duration balloon-borne experiment with
the detectors located on the top of the balloon rather than in a
gondola hanging below the balloon. 
The extended observation time ($\sim$2 weeks) made possible by LDB will permit 
a substantial fraction of the flight to be dedicated to studying and 
characterizing systematics in-flight while still maintaining high
sensitivity to CMB anisotropy.
TopHat will observe in five
spectral bands between 5 and 21 cm$^{-1}$ (150 and 630 GHz
or wavelengths between 0.5 to 2.0 mm).
The current plans call for the measurement of 40 points on the sky,
each with an rms sensitivity of $\delta T_{rms} \approx 1~\mu$K
or $\delta T_{rms} / T_{CMB} \approx 3 \times 10^{-7}$
including removal of the galactic foreground dust emission.

\subsection{ACE/BEAST}
As a follow up to their South Pole HEMT observations the Santa Barbara
group has proposed ACE (Advanced Cosmic Explorer). It is a large,
light-weight (200 kg), system aimed at making flights lasting 90 days or more.
They plan to utilize advanced HEMTs, active refrigerators, and a 2-m diameter
mirror to cover the frequency range 25 to 90 GHz.
In three such flights such a system could map 75\%\ of the sky
to an angular resolution of 10 arcminutes at a level of about 20 $\mu$K.
This project is still in the early phase but is indicative of what
with sufficient funding one might achieve by the year 2000.

\subsection{Ground-Based Instruments}
Ground-based instruments have made a significant contribution to
CMB anisotropy observations. They have been successful 
as a result of the observers' clever strategies
to minimize and reduce the effect of the atmosphere.
These strategies have included going to high, dry sites such as
the South Pole and Teide peak on Tenerife and using triple-beam chopping
or other similar techniques. These techniques are more difficult
to use when going to mapping and making observations
over an extended portion of the power spectrum.
Here again it is possible that significant progress can be made
though it is likely to be eventually limited before the
science is exhausted.

An exciting exception is the use of aperture synthesis interferometers.
The Ryle Telescope images of the Sunyaev-Zeldovich effect in clusters
and the CAT (Cambridge Anisotropy Telescope) results have convinced
many that interferometers have a bright future in actually mapping
anisotropy on small angular scales over selected regions of the sky.

\subsubsection{CAT: Cambridge Anisotropy Telescope; 30$^\prime$ to 2$^\circ$}
The CAT \cite{Robson93} 
is a three-element interferometer 
which can operate at frequencies between 13 and 17 GHz 
with a bandwidth of 500 MHz.
This frequency range was chosen as a compromise between the effects 
of atmospheric emission, which increase with frequency, 
and Galactic synchrotron and bremsstrahlung emission, 
which decrease with frequency. 
The most important contaminating signal for the CAT is
that from discrete extragalactic radio sources.
The observation strategy is to chose fields with minimum source content
and then observe the sources with the higher resolution Ryle Telescope
at 15.7 GHz.

The CAT has a system temperature of approximately 50 K. 
Variations in the system temperature are continuously measured using a 
modulated 1-K noise signal injected into each antenna. 
The interferometer baselines can be varied from 1 to 5 m, 
and are scaled to give the same synthesized beam at different frequencies. 
The antennas have a primary beam FWHM of 2.2 degrees at 15 GHz. 
The CAT simultaneously records data from the two orthogonal linear polarizations. 
Its alt-az mount causes the plane of polarization to rotate on the sky 
as the telescope tracks a given field. 

The CAT is situated within a 5-m high earth bank which is lined with aluminium. 
This shielding reduces the effect of spillover and terrestrial radio
interference, but limits observations to elevations above 25 degrees.
The control hut is located about 100 m away. 
Each element of the telescope is a corrugated-conical horn with a parabolic 
reflector. 
The horns are mounted on a single turntable which can track in azimuth. 
Each antenna has an individual elevation drive. 
Preliminary test have shown that crosstalk, correlator offsets, and antenna 
shadowing - particular problems associated with interferometers - do not
affect the performance of the CAT at elevations greater than 40 degrees
\cite{Robson94}, \cite{OSullivan95}. 
Results \cite{Hobson95} are shown in the summary figure of anisotropies.

\subsubsection{Interferometers: VSA, CBI, \& VCA}
Three major interferometer projects are funded and underway. 
They are the VSA (Very Small Array, 15$^\prime$ to 4$^\circ$) in England, 
the Caltech interferometer CBI (Cosmic Background Interferometer, 
4$^\prime$ to 20$^\prime$) and
the University of Chicago VCA (Very Compact Array, 15$^\prime$ to 1.4$^\circ$). 
These interferometers are likely to provide a very good first
look at the CMB anisotropy power spectrum on angular scales less
than about 0.5$^\circ$ ($\ell \gt 200$).

\begin{table}[t]
\caption{Characteristics of Next Generation Interferometers}
\vspace{0.4cm}
\begin{center}
\begin{tabular}{|c|c|c|c|}
\hline & & & \\ 
Initials: & VSA & VCA & CBI \\
\hline
Name: & Very & Very & Cosmic \\
& Small & Compact & Background \\
& Array & Array & Imager \\
\hline 
Frequency (GHz) & 28 - 37 & 26 - 36 & 26 - 36 \\ 
N$_a$ & 14, 15  & 13 & 13  \\ 
No. of Channels & 1 tunable & 10  & 10 \\ 
T$_{sys}$ (K) & $\sim$25 & $\sim$25 & $\sim$25 \\ 
$\Delta \nu$ & 1.75 & 1 & 1 \\ 
${\ell}$ range & 150-1600 & 150-750 & 400-3500 \\ 
Resolution & 15$^\prime$ & 15$^\prime$ & 4.5$^\prime$ \\ 
Site & Tenerife & Anarctica & California, Chile \\ 
Point Sources & Ryle \& Bonn & A. T. & VLA \& 40-m \\ 
Correlations & analog & analog & analog \\ 
Operational & 1999 & 1999 & 1999 \\ 
\hline 
\end{tabular} 
\end{center} 
\end{table}

VSA is a joint project between the Mullard Radio Astronomy Observatory
(Cambridge), the Nuffield Radio Astronomy Laboratories 
(Jodrell Bank, Manchester),
and the Instituto de Astrofisical de Canarias (Tenerife).
The VSA is similar in design to CAT which was a prototype for VSA.
The VSA will have 15 antennas and a 2-GHz bandwidth, analog correlators,
and other technology operated on CAT.
The operating frequency is 26-36 GHz which is set by the atmospheric
window and the natural waveguide bands for which high sensitivity HEMT 
amplifiers have been developed.
The increase in frequency from CAT to VSA will also decrease the effect
of discrete radio sources and Galactic emission.
The atmospheric emission fluctuations will increase so that
VSA will be operated on Mt. Teide on the island of Tenerife.
The VSA will operate with two sets of horns: 
one set with a 15-cm aperture giving a 4$^\circ$ field of view
and the second with a 30-cm aperture giving a 2$^\circ$ field of view.
The baselines and thus resolution will scale proportionally
to maintain about a 1 $\mu$K sensitivity per resolution element.
The VSA will get about 10 independent points of the anisotropy power spectrum
with resolution of $\Delta \ell = 100$ at low $\ell$ and
$\Delta \ell = 200$ at high $\ell$ covering the range $ 130 < \ell < 1800$.

The CBI and VCA are planning to observe the same portion of sky
in the southern hemisphere.
The proposed VCA is expected to image
about 2500 square degrees around the South Pole region
and if that goes well continued operation to cover eventually 
about 25\%\ of the sky.
The VCA interferometer consists of 13 scalar feed horns
arranged in a closed packed configuration which fill about 50\%\ 
of the aperture to provide maximum brightness sensitivity.
The horns feed low-noise HEMT amplifiers operating at 26 to 36 GHz
with noise temperatures of about 10 K. 
The estimated sensitivity is 4 to 10 $\mu$K in pixels
ranging from 0.25$^\circ$ to 1.4$^\circ$ in the 3$^\circ$ field of view.
The VCA will be operated from the South Pole Station and
is scheduled for installation in the fall of 1998 
with first results expected the following spring.

The sensitivity of an interferometer system can be estimated using
the following formulae for flux density and temperature:

Flux density
$$
\Delta S_{rms} = \frac{2 k T_{sys} }
{ \eta_a A_a \eta_c [ n_a (n_a -1) \Delta \nu \tau ]^{1/2} }
$$

Temperature
$$
\Delta T_{rms} = \frac{ \lambda^2 T_{sys} }
{\theta_s^2 \eta_a A_a \eta_c [ n_a (n_a -1) \Delta \nu \tau ]^{1/2} }
$$

Same expressions evaluated with typical numbers:

Flux density
$$
\Delta S_{rms} = \frac{6 (\frac{ T_{sys}} {30~K} ) } 
{ (\frac{\eta_a} {0.6}) ( \frac{d}{20~cm} )^2 ( \frac{\eta_c}{0.9} )
[ \frac{n_a}{14} \frac{n_a - 1}{13} \frac{\Delta \nu}{10^9} 
\frac{\tau}{1~month} ]^{1/2} }
{\rm mJy}
$$

Temperature
$$
\Delta T_{rms} = \frac{ 6 (\frac{\lambda}{1~cm} )^2 (\frac{T_{sys}}{30~K}) }
{ ( \frac{\theta_s} {20^\prime} )^2
(\frac{\eta_a} {0.6}) ( \frac{d}{20~cm} )^2 ( \frac{\eta_c}{0.9} )
[ \frac{n_a}{14} \frac{n_a - 1}{13} \frac{\Delta \nu}{10^9}
\frac{\tau}{1~month} ]^{1/2} }
\mu K
$$

One can then evaluate these formula and compare with the table
to estimate the sensitivity and sky area that can be surveyed
in a given observing time and see that on small angular scales 
interferometers are competive with many other experiments.

\section{Future Satellite Missions}
An accurate, extensive imaging of CBR anisotropies
with sub--degree angular resolution would provide
decisive answers to several major open questions
on structure formation and cosmological scenarios.
The observational requirements of such an ambitious objective
can be met by a space mission 
with a far--Earth orbit and instruments based on state--of--the--art
technologies.  

While balloon-borne and ground-based observations can do a credible
job in measuring the CMB anisotropy power spectrum,
atmospheric disturbance, emission from the Earth and limited
integration time are the main limiting factors
which prevent ground--based and balloon--borne experiments
from obtaining sufficient sensitivity over very large sky regions,
with additional difficulties in reaching accurate foreground
removal (see Danese et al. 1995 for a recent discussion).
Only a suitably designed space mission can meet the scientific
goals sought by cosmologists.
On the other hand it should be stressed that experiments
from the ground or from balloons
are not alternative to a space mission like Planck,
but rather complementary. 

\subsection{{\bf MAP}}
MAP (Microwave Anisotropy Probe) has been selected by NASA in 1996 as
a MidEX class mission. Its launch is expected to be roughly 2001.
The goal of MAP is to measure the relative CMB temperature over the full sky 
with an angular resolution of 0.3$^\circ$, a sensitivity of 20 $\mu$K 
per 0.3$^\circ$ square pixel, and with systematic effects limited to 5 $\mu$K 
per pixel. Details about the major aspects of the mission design are given
below. 

\subsubsection{Galactic Emission Foreground}
Galactic foreground signals are distinguishable from CMB anisotropy by their 
differing spectra and spatial distributions. 

\begin{figure}
\centerline{\epsfxsize= 8 cm \epsfbox[0 0 612 792]{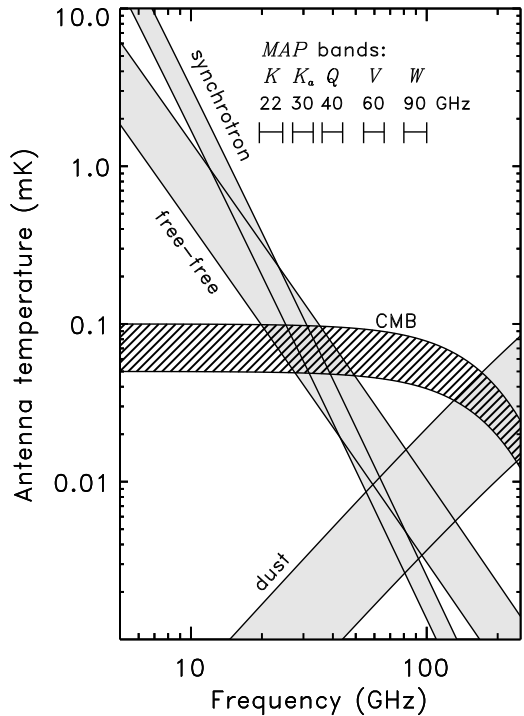}}
\caption
{Galactic Foreground Emission estimates plotted as a function of frequency.
The width of the band indicates the estimated range for Galactic latitudes
varying between  $20^\circ < b < 70^\circ$.
The proposed MAP frequency bands are indicated.
}
\label{MAPG}
\end{figure}

Figure \ref{MAPG} shows the estimated spectra of the
Galactic foreground signals and a range of expected
cosmological signal intensities. The three physical
mechanisms that contribute to the Galactic emission
are synchrotron radiation, free-free radiation, and
thermal radiation from interstellar dust. Results
from CMB and other measurements show that at
high Galactic latitudes CMB anisotropy dominates
the Galactic signals in the range ~30-150 GHz.
However, the Galactic foreground will need to be
measured and removed from some of the MAP data. 

There are two techniques that MAP will use to
evaluate and remove the Galactic foreground. The
first uses existing Galactic maps at lower (radio)
and higher (far-infrared) frequencies as foreground
emission templates. Uncertainties in the original
data and position-dependent spectral index
variations introduce errors with this technique.
There is no good free-free emission template
because there is no frequency where it dominates
the microwave emission. High resolution, large
scale maps of H-alpha emission will be a template
for the free-free emission when they become available. 
One surprise has been the apparent correlation between free-free emission
and dust infrared emission, sometimes more so than to H-alpha emission.
This indicates that we may not yet have a template for free-free emission.

The second technique is to form linear combinations of multi-frequency MAP 
observations such that signals with specified spectra are cancelled. 
The linear combination of multi-frequency data makes no assumptions 
about the foreground signal strength or spatial distribution, but requires
knowledge of the spectra of the foregrounds. 
Both techniques were successfully employed by COBE. 

The contamination from extragalactic radio sources is not yet
a solved issue.
Flat spectrum radio sources observed with a diffraction limited system
produce a signal with very nearly the same frequency dependence
as the CMB anisotropy making the spectral shape technique ineffective.
It will be necessary to compile a list of significant radio sources
and excise them from the data or find another approach.

Five frequency bands with comparable sensitivity are desirable to solve 
for the four signals (synchrotron, free-free, dust, and CMB anisotropy) 
and the fifth degree of freedom is used to maximize signal-to-noise. 
The range of frequency coverage is more important than the specific
choice of frequencies within the range. 
The lowest frequency to survey from space should be at the
22 GHz atmospheric water line since frequencies below this can 
(with difficulty) be accurately measured from the ground. 
The highest frequency to survey should be about 100 GHz to reduce the dust
contribution and minimize the number of competing foreground signals. 
The choice of frequencies between 22 and 100 GHz can be dictated by the 
practical consideration of standard waveguide bands. 
Based on these considerations, MAP has selected the five frequency 
bands, which are indicated Figure 9 and in Table 2.

\subsubsection{MAP Mission Goals}
CMB anisotropy information from current and proposed high resolution 
($< 0.3^\circ$) measurements over limited sky regions will likely succeed 
from ground and balloon-based platforms. 
The priority for the MAP mission is to map the entire sky with $> 0.3^\circ$ 
angular resolution where the cosmological return is high, 
and the data cannot be readily obtained in any other way. 
The MAP optics feature back to back 1.5-meter primary reflectors 
which lead to an angular resolution of 0.29$^\circ$ in the highest frequency 
(90 GHz) channel. 

The following table gives the angular resolution to be obtained from each 
of the five MAP frequency bands. The value quoted is the full width 
at half maximum (FWHM) of the approximately Gaussian central beam lobe, 
in degrees. 

\begin{table}[htb]
\begin{center}
\caption{MAP Angular Resolution}
\begin{tabular}{lcccccc}
\hline 
Frequency & Band    & Wavelength & Beam &FWHM & No. of& Sensitivity \\
 (GHz)   &  Name    &    (mm)    & Spec &Design & Channels & 0.3$^\circ$ by 0.3$^\circ$ pixel    \\
\hline
22 GHz & K band & 13.6 & 0.90$^\circ$ & 0.93$^\circ$ & 4 & 35$\mu$K 26$\mu$K \\
30 GHz & Ka band& 10.0 & 0.65$^\circ$ & 0.68$^\circ$ & 4 & 35$\mu$K 32$\mu$K \\
40 GHz & Q band &  7.5 & 0.53$^\circ$ & 0.47$^\circ$ & 8 & 35$\mu$K 27$\mu$K \\
60 GHz & V band &  5.0 & 0.39$^\circ$ & 0.35$^\circ$ & 8 & 35$\mu$K 35$\mu$K \\
90 GHz & W band &  3.3 & 0.29$^\circ$ & 0.21$^\circ$ &16 & 35$\mu$K 35$\mu$K \\
\hline
\end{tabular}
\end{center}
\end{table}

The MAP specification calls for an equal noise sensitivity per frequency band 
of 35 $\mu$K per $0.3^\circ \times 0.3^\circ$ square pixel. 
The mission duration required to meet this specification is one year of
continuous observation. 
If Galactic emission is negligible at high latitudes above 40 GHz, 
as was the case for COBE, the sensitivity achievable by combining 
the three highest frequency channels is ~20 $\mu$K per 
$0.3^\circ \times 0.3^\circ$ pixel. 

The corresponding sensitivity to the angular power spectrum, obtained with 
simple analytic formulae, is illustrated in Figure 10 which shows 
the predicted power spectra for a number of competing structure formation 
models. The gray band straddling the solid (CDM) curve indicates
the MAP sensitivity after combining the three highest frequency channels and 
averaging the spectrum over a 10\%\ band in spherical harmonic order.

\begin{figure}
\centerline{\epsfxsize= 9 cm \epsfbox[0 250 702 602]{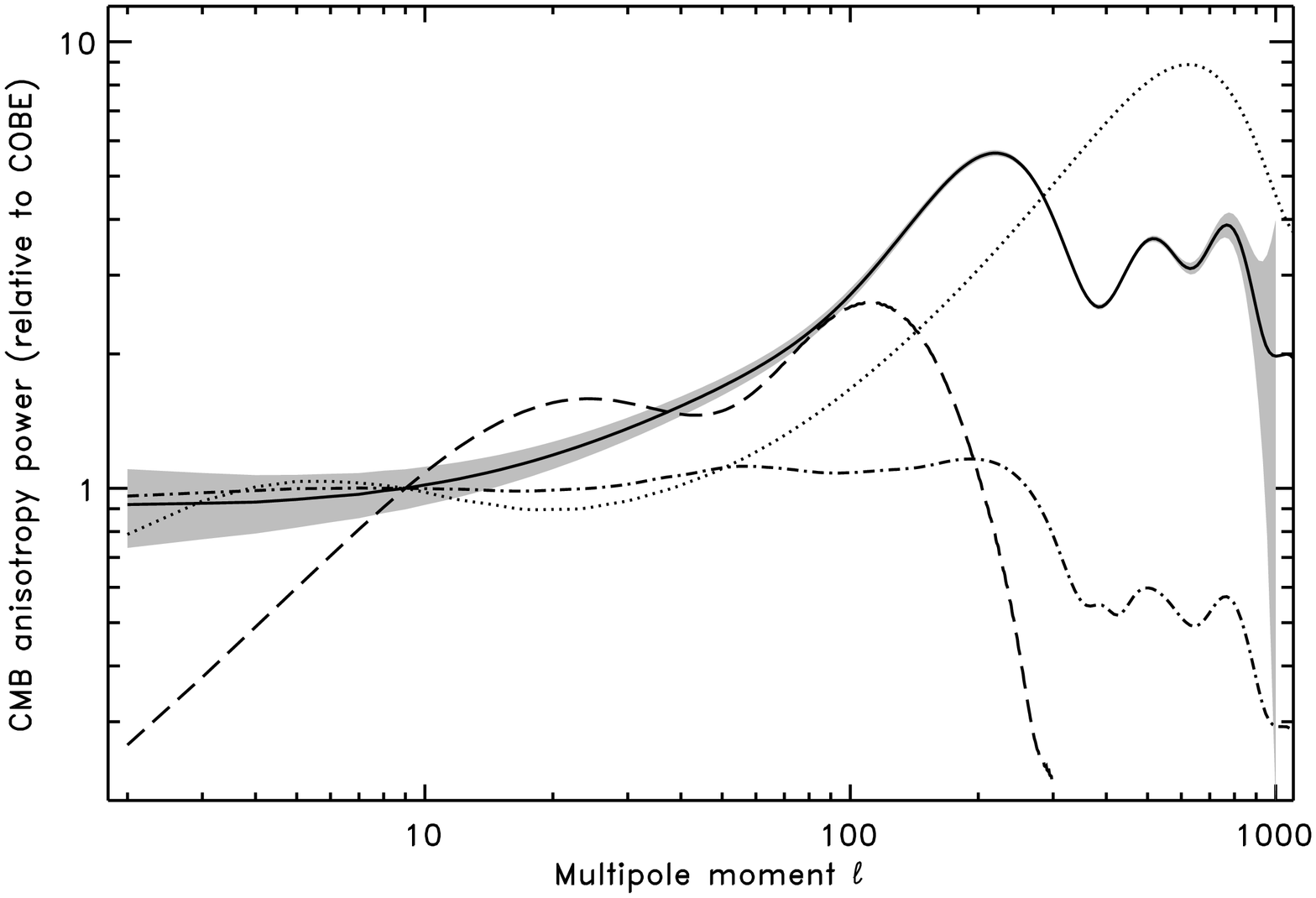}}
\caption
{Projected MAP Power Spectrum Sensitivity.
This plot shows a simple-minded estimate of the MAP sensitivity in 
measuring the CMB power spectrum.
The gray band straddling the solid (CDM) curve indicates
the MAP sensitivity after combining the three highest frequency channels and 
averaging the spectrum over a 10\%\ band in spherical harmonic order.
The curves plotted correspond to a standard CDM model (solid), 
a highly reionized CDM model (dot-dash), 
an open CDM model (dotted), and a primordial baryon isocurvature (PBI) model
(dashed). 
}
\label{MAPcl}
\end{figure}

\subsubsection{MAP Trajectory and Orbit}

To minimize environmental disturbances and maximize observing efficiency, MAP
will observe from a Lissajous orbit about the L2 Sun-Earth Lagrange point 1.5
million km from Earth. 
The trajectory selected to attain such an orbit consists of
2.5-3.5 lunar phasings loops followed by about a 100 day cruise to L2. 
No thruster firings are required to enter the L2 orbit. 

The L2 Lagrange point offers a virtually ideal location from which to carry out 
CMB observations. Because of its distance, 1.5 million km from Earth, 
it affords great protection from the Earth's microwave emission, 
magnetic fields, and other disturbances. 
It also provides for a very stable thermal environment and near 100\%\ 
observing efficiency since the Sun, Earth, and Moon are
always behind the instrument's field of view. 

The following description indicates the path MAP will follow to L2. 
The trajectory features 2.5 or 3.5 lunar phasing loops which assist the 
spacecraft in reaching L2. The cruise time to L2 is approximately 100 days 
after the lunar phasing loops are completed. The launch window for this
trajectory is about 20 minutes/day for 7 consecutive days each month. 
Once in orbit about L2, the satellite maintains a Lissajous orbit such 
that the MAP-Earth vector remains between 1$^\circ$ and 10$^\circ$ 
off the Sun-Earth vector to satisfy communications requirements 
while avoiding eclipses.
Station-keeping maneuvers will be required about 4 times per year to maintain 
this orbit. 

\subsubsection{MAP Instrumentation}
The MAP instrument consists of two back to back, off axis Gregorian telescopes
that produce two focal planes, A and B, on opposite sides of the spacecraft
symmetry axis. 
A set of 10 corrugated feeds lie in each focal plane and collect the signal
power that goes to the amplification electronics. 
The microwave system consists of 10 4-channel differencing assemblies 
that are designed to eliminate low frequency
gain instabilities and amplifier noise in the differential signal. 

\begin{figure}
\centerline{\epsfxsize= 14 cm \epsfbox[ 0 0 625 864]{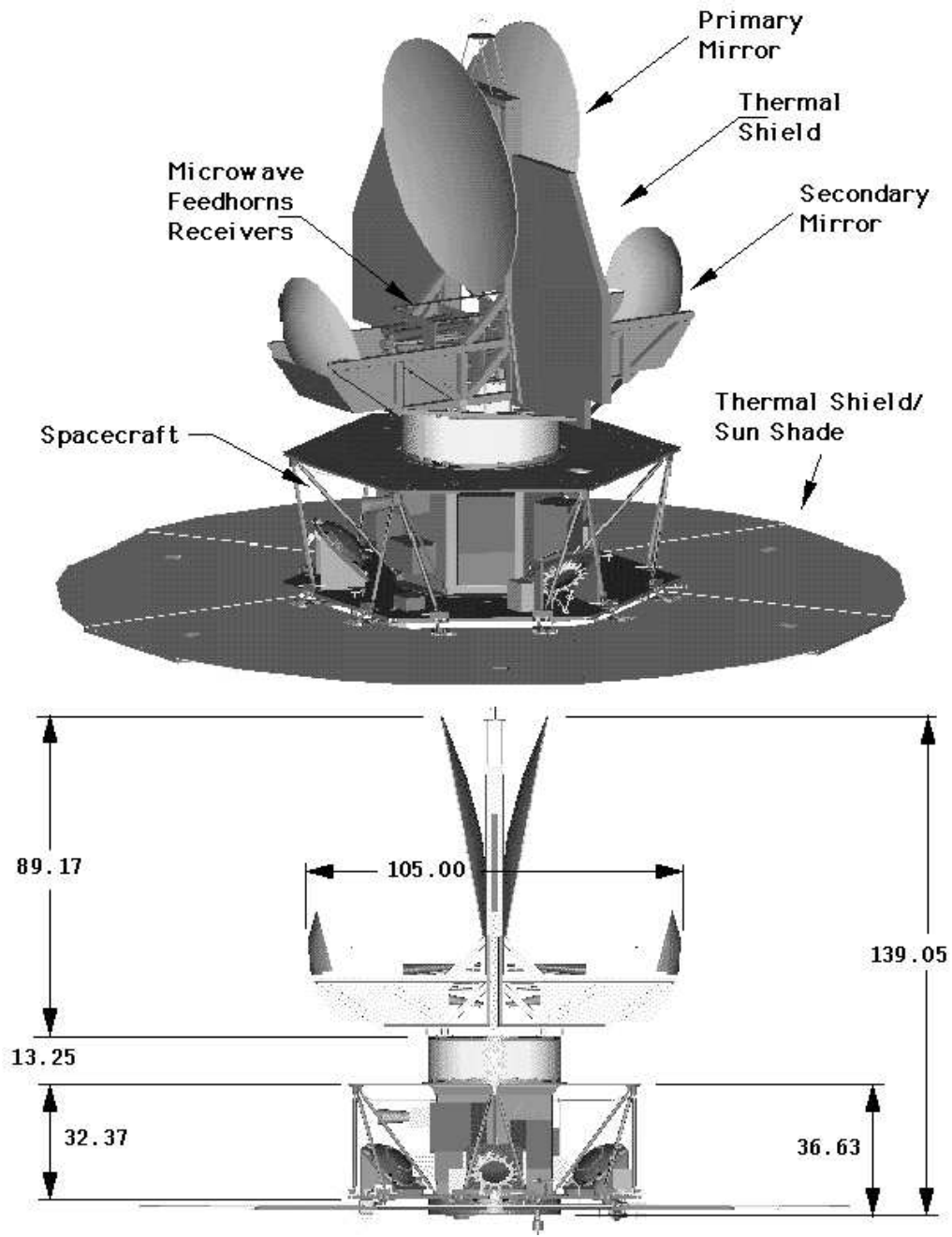}}
\caption
{MAP (Microwave Anisotropy Probe) artists conception shown in two views.
}
\label{MAPdual}
\end{figure}

The reflector design incorporates two back-to-back off-axis Gregorian 
telescopes with 1.5-m primary reflectors and 0.52 m secondary reflectors. 
Each primary is an elliptical section of a paraboloid, while the secondaries 
are nearly elliptical. This arrangement produces two slightly
convex focal surfaces on opposite sides of the spacecraft spin (symmetry) axis 
with plate scales of  about 15$^\prime$/cm. 
The 99.5\%\ encircled energy spot size diameter is less than 1 cm over a 
15 x 15 cm region of the focal plane, and less than 0.33 cm over the central 
8 x 8 cm region. 

In order to limit diffracted signals to less than 0.5 $\mu$K, 
diffraction shields are employed above, below, and to the sides of each 
secondary. In addition, the deployable solar panels and multi-layer
insulation guarantee that the secondaries remain at least 6$^\circ$ 
into the shadow from the Sun during observing. 

The feed design calls for as small an aperture as possible consistent 
with a primary edge taper requirement of -25 dB, and a length 
that places the throat of each differential feed pair in close
proximity to the other. 
The feed aperture diameters scale inversely with frequency, while the
primary is equally illuminated at each frequency, leading 
to a frequency dependent beam size. 
The feeds are corrugated to produce beams with high symmetry, low-loss, 
and minimal sidelobes: the extremely low loss HE$_{11}$ hybrid mode dominates. 
The phase center of each feed is kept as close as possible to its aperture, 
resulting in a frequency-independent beam for each feed. 
Since the distance from the focal plane to the spacecraft symmetry axis is 
nearly the same for all the feeds, the high frequency feeds are extended 
with low loss corrugated waveguide, while the low frequency
feeds are ``profiled" to reduce their length, 
while limiting excitation of the TE$_{11}$ mode to less than -30 dB. 

The microwave system consists of 10 4-channel differencing assemblies, 
one for each pair of feeds. 
One assembly operates at 22 GHz, one at 30 GHz, two at 40 GHz, two at 60 GHz, 
and four at 90 GHz. 
The base of an A-side feed in the Focal Plane Assembly (FPA) is attached 
to a low-loss orthomode transducer (OMT) which separates the signal 
into two orthogonal polarizations, A and A$^\prime$. 
The A side signal is differenced against the orthogonal polarization, 
B$^\prime$, from the corresponding B-side feed, and vice-versa. 

The differencing is accomplished by first combining the two signals A and 
B$^\prime$  
in a hybrid tee to form (A+B$^\prime$)/$\sqrt{2}$ and (A-B$^\prime$)/$\sqrt{2}$, 
then amplifying each in two cold HEMT amplifiers and
sending the phase-matched outputs to the warm receiver box via waveguide. 
The two signals are amplified in two warm HEMT amplifiers, 
phase switched between 0$^\circ$ and +90$^\circ$ or -90$^\circ$, respectively, 
at 2.5 kHz, then split back into A and B$^\prime$ in a second hybrid tee.
At this point, the two signals are square-law detected, amplified 
by two line drivers, and sent to the Analog Electronics Unit for synchronous 
demodulation and digitization. The other pair of signals,
A$^\prime$ and B, are differenced in the same manner giving a total 
of four amplification channels per differencing assembly. 

The splitting, phase switching, and subsequent combining of the signals enhances 
the instrument's performance in two ways: 

(1) Since both signals to be differenced are amplified by both amplifier chains, 
gain fluctuations in either amplifier chain act identically on both signals 
and thus cancel upon differencing. 
(2) The phase switches introduce a 180 degree relative phase change between the two 
signal paths, thereby interchanging which signal is fed to which square law 
detector. Thus, low frequency noise from the detector diodes is common mode 
and also cancels, further reducing susceptibility to systematic effects. 

\subsubsection{Map Making with Differential Data}

MAP will observe temperature differences between points separated by 135$^\circ$
on the sky. Maps of the relative sky temperature will be produced 
from the difference data by a modification of the algorithm used by
COBE-DMR. 

The algorithm MAP will use to reconstruct sky maps from differential data is 
iterative. It is mathematically equivalent to a least squares fitting
of the temperature differences to the map pixel temperatures. 
However, the scheme has a very intuitive interpretation: for a given pair of
differential feeds, A and B, the A feed can be thought of as viewing the sky 
while the B feed can be thought of as viewing a comparative reference signal, 
or vice versa. 
In MAP's case, the comparative signal is a different point in the sky. 
The actual signal MAP measures is the temperature difference 
between two points on the sky, $\Delta$T = T(A)-T(B), 
where T(A) is the temperature seen by feed A, and likewise for B. 
If the temperature T(B) is known, one could recover T(A) 
using T(A) = $\Delta$T+T(B), but since T(B) is not known, 
the algorithm makes use of an iterative scheme 
in which T(B) is estimated from the previous sky map iteration. 
Thus the temperature in a pixel of a map is given by the average 
of all observations of that pixel after correcting each observation 
for the estimated signal seen by the opposite feed. 

For this scheme to be successful it is imperative for a given pixel to be 
observed with many different pixels on its ring of 135$^\circ$ partners away.
Thus the method requires a carefully designed scan strategy. 
The MAP strategy achieves this while simultaneously avoiding
close encounters with the Sun, Earth, and Moon. 
The algorithm has been tested with the MAP scan strategy using an end-to-end 
mission simulation that incorporates a realistic sky signal, instrument
noise, and calibration methods. The results of these simulations are described 
in detail in an Astrophysical Journal article \cite{Wright96b}.
After 40 iterations of the algorithm, the artifacts that remain in the map due 
to the map-making itself have a peak-peak amplitude of less than 0.2 $\mu$K, 
even in the presence of Galactic features with a peak
brightness in excess of 60 mK. 

\subsubsection{MAP Sky Coverage and Scan Strategy} 

MAP will observe the full sky every six months. The MAP scan strategy 
combines spacecraft spin and precession to achieve the following: 
1) The MAP instrument observes more than 30\%\ of the sky each day; 
2) The spacecraft spin (and symmetry) axis maintains a fixed angle of 
22.5$^\circ$ from the Sun-Earth line to mitigate systematic effects; and 
3) Each sky pixel is connected to thousands of other sky pixels to ensure 
high quality map solutions with negligible noise correlations. 

\begin{figure}
\epsfysize=4.0truein
\epsfverbosetrue
\centerline{\epsfbox[7 157 604 634]{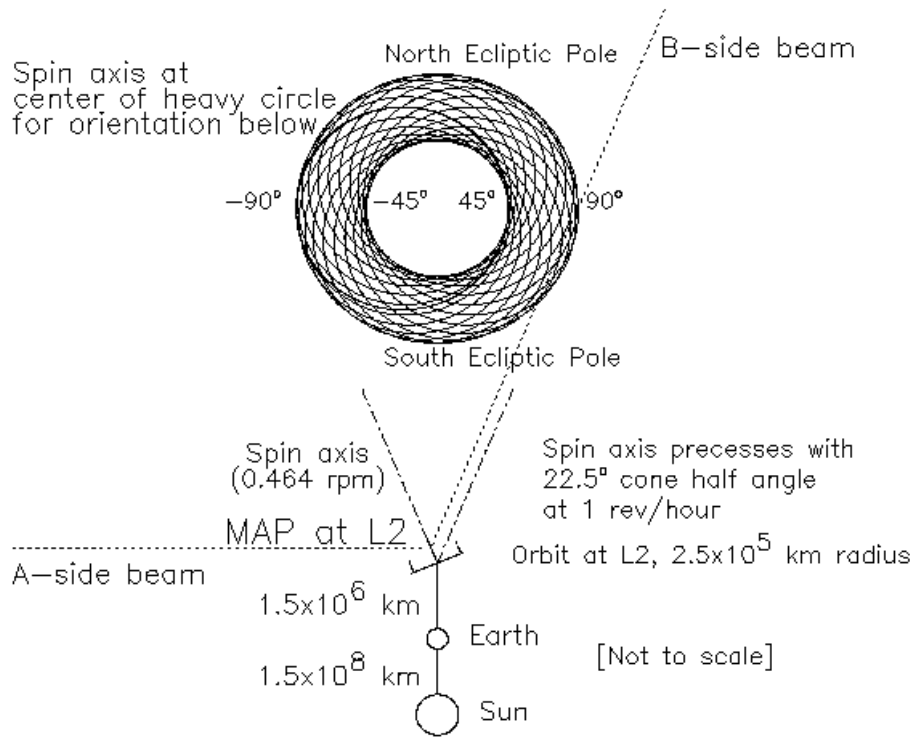}} 
\caption{
The MAP scan pattern for one hour of observation.
The lines show the path for one side of a differential antenna pair. 
The other pair member follows a similar path, only delayed by 1.1 min. 
There are four principal time scales for the observations.
The phase of the difference signal is switched by $180^{\circ}$ at 2.5 KHz. 
The spacecraft spins around its symmetry axis with a 2.2 min period 
(bold circle) with cone opening angle of roughly $135^{\circ}$. 
This pattern precesses about the Earth-Sun line with a period of 60 minutes. 
Thus, in about 1 hour, over 30\% of the sky is covered. 
Every six months, the whole sky is observed.
Note that any pixel is differenced to another pixel in many directions. }
\end{figure}

Since a major goal of cosmology is to determine the statistical properties 
of the universe, it is clear that the largest possible number of sky samples 
improves constraints on cosmological models. 
The measurement of each individual position on the sky is an independent sample 
of the cosmology of the universe. Moreover, full sky coverage is absolutely 
required to accurately determine the low-order spherical harmonic moments. 
While the largest angular scales were observed by COBE,
MAP will remeasure the full sky with higher resolution to: 

$\bullet$ Avoid relative calibration errors when two or more experimental results are 
combined (eg., COBE and MAP). 

$\bullet$ Provide greater sensitivity to the angular power spectrum. 

$\bullet$ Independently verify the COBE results. 

The goals of the MAP scan strategy include the following: 

The angular separation between the two observing beams should be ``large" in
order to maintain sensitivity to signal at large angular scales. 
This is important for comparing the MAP results to COBE, for properly 
normalizing the angular power spectrum, and for retaining sensitivity 
to the dipole which will serve as MAP's primary calibration source. 

$\bullet$ Observe a large fraction of the sky every day. 
This guarantees that sky pixels will be observed at many different times 
in the mission which provides the capability to monitor instrument stability 
on many different time scales. 

$\bullet$ Maintain a fixed angle between the spacecraft spin axis and 
the local solar vector. 
This provides for stable illumination of the spacecraft solar panels which lie 
normal to the spin axis, and provides a thermally stable environment to 
mitigate systematic effects. 

$\bullet$ Connect each sky pixel to as many other sky pixels as possible 
to provide high quality map solutions from the differential data, 
and to render negligible pixel-pixel noise correlations. 
Since the MAP beam separation is fixed, this implies observing as many pixels 
on the differential ring of pixels as possible. 

The MAP beam separation is 135$^\circ$. 
Each beam axis points 67.5$^\circ$ away from the spin and symmetry axis 
of the spacecraft. 
The spin axis will precess in a 22.5$^\circ$ angle about the local solar vector. 
The combined spacecraft spin and precession will cause the observing beams
to fill an annulus centered on the local solar vector with inner and 
outer radii of 45$^\circ$ and 90$^\circ$ respectively. 
Thus MAP will observe more than 30\%\ of the sky each day and will observe the
ecliptic poles every day. 
The spin period will be 2.2 minutes while the precession period will be 1 hour. 
As the Earth orbits the Sun, the whole observing annulus revolves 
with it producing full sky coverage.

The MAP mission is moving ahead quickly. Its spacecraft preliminary design
review occurred in January 1997 and instrument held in March 1997.
At that point many of the major design features were
fixed and only smaller modifications will occur.
However, MAP has a WWW page http://map.gsfc.nasa.gov
which can be consulted for the latest information.

\subsection{{\bf Planck -- The Third Generation Space Mission}}
The Planck mission is the result of the
merging of two proposals presented in 1993 to the
European Space Agency {\it M3 Call for
Mission Ideas}: COBRAS (Cosmic Background Radiation Anisotropy
Satellite \cite{Mandolesi93}
and SAMBA (Satellite for Measurements of Background Anisotropies
\cite{Puget93}). The COBRAS/SAMBA team completed the ESA
assessment study in May 1994, and the project continued and
completed the Phase A study in May 1996. 
COBRAS/SAMBA, renamed Planck Surveyor,
has been selected to continue within the European Space Agency M3 programme.

The Planck mission is designed for extensive, accurate
mapping of the anisotropy of the CMB, 
with angular sensitivity from sub--degree ($\sim 8^\prime - 30^\prime$)
scales up to the full sky thus overlapping with the COBE--DMR maps
and with signal sensitivity approaching $\Delta T/T \sim 10^{-6}$.
Planck will survey the entire sky at frequencies from 30 to 850\,GHz
($1\,{\rm cm} \leq \lambda \leq 350$~microns).   
Its 1.3-m passively cooled telescope will be diffraction limited 
at frequencies below 375\,GHz.   
The primary science goal is a definitive measurement of the structure 
in the CMB on all angular scales of 10$^\prime$ 
or larger.   The case for orbital measurements of CMB anisotropy has been well
made.  An accurate measurement of CMB anisotropy with angular resolution of
10$^\prime$ will revolutionize cosmology.   
A full description of the baseline Planck mission is available from 
ESTEC at http://astro.estec.esa.nl/SA-general/Projects/ COBRAS/cobras.html.  
Additional information is available at
http://aether.lbl.gov/www/cosa/ 
This information will be updated
regularly and new links added.

Planck has been selected as ESA's next medium-scale mission (M3), 
this was confirmed in the November 1996 review of the Ariane V demise 
of the Cluster Mission.
Planck is scheduled to fly in 2004,  
approximately four years after the MAP NASA Midex mission. 
Planck is well-designed to follow \hbox{MAP}.  
Using more sophisticated detector technology, Planck
will have 10 times the sensitivity, 2 or 3 (depending on frequency) times the
angular resolution, and 6 times the frequency coverage of MAP.  
This performance will allow Planck to:

\smallskip\noindent 1. Measure the power spectrum of the CMB with accuracy
limited by cosmic variance over almost the entire range of angular frequency
space in which useful cosmological information is expected.

\smallskip\noindent 2. Separate Galactic and extragalactic foregrounds from the
CMB with high accuracy and confidence.  The broad frequency coverage will allow
determination of all important foreground components without any prior
assumptions about their spectra.

\smallskip\noindent 3. Separate secondary anisotropies due to the
Sunyaev-Zel'dovich effect from primary anisotropies, and measure the SZ effect
with precision in thousands of clusters of galaxies.  This information,
combined with X-ray data, will  yield an independent measurement of the Hubble
constant on large scales and probe the peculiar velocity field of clusters to
high redshift.

\smallskip\noindent 4. Separate polarization of the CMB from that in local
foregrounds and  measure it with precision on angular scales as small as 
7$^\prime$.

\smallskip\noindent
5. Survey the sky at sub-mm (350, 550 and 850 microns) wavelengths
that complement the wavelength coverage of SIRTF.  The chance for
serendiptous discovery in this survey is great.

Planck has two instruments: the Low Frequency Instrument (LFI), based on
transistor (HEMT) amplifiers, which covers the frequency range from 30--100\,GHz; and
the High Frequency Instrument (HFI), based on bolometers, which covers the
frequency range 90--850\,GHz. 
The implementation of these technologies fits comfortably within the mass, 
power, volume and schedule constraints of the ESA M3 opportunity.

\subsubsection{Planck Active Cooling Option}
One area being pursued at the present is the use of actively
cooled HEMTs, which requires the HEMT amplifier chain to be broken into
a low-temperature portion and a higher-temperature portion.
This split reduces the thermal load on the focal plane allowing
passive cooling to a significantly lower temperature (i.e., 65~K)
and allowing for the use of active cooling technologies such as
a hydrogen sorption cooler.

In addition, there are technical advantages of a combined LFI/HFI focal assembly
that uses a hydrogen sorption cooler to cool both the HEMTs and the
20\,K shield of the bolometer dewar. The sorption cooler
would replace the 20\,K Stirling cooler in the baseline design, with its
attendant problems in vibration and instrument integration, and reduce the
overall mass and power of the instruments.  This is an extremely attractive
option that will be studied in detail by both the HFI and LFI teams.

The split HEMT design and sorption cooler enable a LFI design with the
following features:

$\bullet$ An increase in the sensitivity of the LFI by a factor of roughly five at
the highest frequency over that of the Phase A baseline design.

$\bullet$ Division of the HEMT radiometers into a cold focal assembly and a
room-temperature assembly.  The power dissipated in the focal assembly is more
than an order of magnitude lower than assumed in the Phase A study design,
allowing radiative cooling of the focal assembly to a temperature of
$\leq65$\,K instead of $\sim100$\,K.

$\bullet$ Active cooling of the HEMTs in the focal assembly to $\leq20$\,K.
This reduces potential thermal interactions between the LFI and HFI.

$\bullet$ Elimination of the $\sim80$\,K Stirling cooler in the baseline design.

$\bullet$ The option (introduced above) of using the sorption cooler to cool the
20\,K shield around the HFI as well as the HEMTs, eliminating the need for the
20\,K Stirling cooler in the baseline design.  An important benefit of this
option is that the vibration associated with Stirling coolers would be
eliminated from the focal assembly.  The overall mass and power requirements of
the instruments on the spacecraft would decrease by roughly 25\,kg and 100\,W
as well.

This design will allow a full identification of the primordial density 
perturbations which grew to form the large--scale structures
observed in the present universe.
The Planck maps will provide decisive answers to several major open
questions relevant to the structure formation epoch
and will provide powerful tests for the inflationary model
as well as several astrophysical issues.
Planck will utilize a combination of bolometric and radiometric detection
techniques to ensure the sensitivity and wide spectral coverage required for
accurate foreground discrimination.
An orbit far from Earth  has been selected to minimize the unwanted emission
from the Earth as a source of contamination.

\subsubsection{Planck Scientific Objectives}

The Planck mission will produce near all-sky maps of the background
anisotropies in 8 frequency bands in the range 30--800 GHz,
with peak sensitivity $\Delta T/T \sim 10^{-6}$.
The maps will provide a detailed description
of the background radiation fluctuations. Individual
hot and cold regions should be identified above the
statistical noise level, at all angular scales from
$\lsim10^\prime$ up to very large scales, thus providing a
high resolution imaging of the last scattering surface.

The Planck maps will provide all multipoles
of the temperature anisotropies from $\ell=1$ (dipole term) up to
$\ell \simeq 1500$ (corresponding to $\sim 7^\prime$). It is the
information contained in this large number of multipoles
that can probe the various proposed scenarios of structure
formation and the shape of the primordial fluctuation spectrum
(for comparison, the COBE--DMR maps are limited to
$\ell \lsim ~20$).

Table~3 compares the ability of various CMB missions to
determine cosmological parameters in a model-dependent but self-consistent
way.  The details of the calculation are not important, with two exceptions. 
First, $\Omega=1$ was assumed.  If $\Omega$ is smaller differences in angular
resolution become even more important.  Second, it was assumed that confusing
foregrounds were completely removed.  In practice this will not be the case. 
The advantage that Planck's wide frequency coverage gives is therefore not
reflected in the table.  Nevertheless, the power of Planck in general, and
the advantages of cooling the HEMTs, are immediately apparent.

\begin{table}[htb]
\begin{center}
\caption{U\sc{NCERTAINTIES IN} C{\sc OSMOLOGICAL} P{\sc ARAMETERS}}
\begin{tabular}{lcccc}
\hline 
\hline 
\omit& I{\sc NSTRUMENT}\\
P{\sc ARAMETER} & MAP & C/S HFI\rlap{$^{\rm a}$}&C/S LFI\rlap{$^{\rm b}$}&
    C/S LFI\rlap{$^{\rm c}$}\\
\hline
$Q_{\rm rms-ps}/20\mu K$ & 0.23 & 0.12 & 0.18 & 0.14 \\
$h$ & 0.13 & 0.032 & 0.12 & 0.065 \\
$h^2\Omega_b$ & 0.0072 & 0.0019 & 0.0062 & 0.0036 \\
$\Omega_\Lambda$ & 0.67 & 0.19 & 0.59 & 0.33 \\
$\Omega_\nu$ & 0.38 & 0.12 & 0.36 & 0.28 \\
$\Omega$ & 0.11 & 0.012 & 0.068 & 0.029 \\
$n_s$ & 0.12 & 0.017 & 0.074 & 0.029 \\
$\tau$ & 0.35 & 0.15 & 0.26 & 0.20 \\
$N_\nu$ & 0.43 & 0.16 & 0.40 & 0.26 \\
$T_0\,\mu K$ & 0.01 & 0.01 & 0.01 & 0.01 \\
$Y$ & 0.01 & 0.098 & 0.01 & 0.01 \\
T/S & 0.47 & 0.17 & 0.28 & 0.19 \\
$Q_{\rm ps}/0.2\mu K$ & 0.02 & 0.000021 & 0.06 & 0.0006 \\
$Q_{\rm diffuse}/20\mu K$ & 0.19 & 0.17 & 0.18 & 0.18 \\
\hline
\end{tabular}
\end{center}

$^{\rm a}$ HFI based on ``spider web" bolometers, as given in Phase A study \\

$^{\rm b}$ LFI based on InP HEMTs at 100\,K, as given in Phase A study \\

$^{\rm c}$ LFI based on InP HEMTs at 20\,K \\
\label{comp}
\end{table}

The high resolution Planck maps will provide
a key test for structure formation mechanisms,
based on the statistics of the observed
$\Delta T/T$ distribution. The
inflationary model predicts Gaussian fluctuations for the
statistics of the CBR anisotropies, while alternative
models based on the presence of topological defects, such as
strings, monopoles, and textures, predict non--Gaussian statistics
(e.g. \cite{Coulson94}). 
Due to the different nature of their early history
causality constrains primordial perturbations from a source
such as inflation and from topological defects to have
a different anisotropy power spectra particularly
in the region of the ``Doppler'' peaks \cite{Albrecht95}.
The angular resolution and sensitivity of Planck
will allow discrimination between these alternatives
with tests of both the power spectrum and statistics.

The high-order multipoles will allow an accurate measure
of the spectral index $n$ of the primordial fluctuation spectrum:
\begin{equation}
(\delta \phi)^2 \propto \lambda^{(1-n)}
\end{equation}
where $\delta \phi$ is the potential fluctuation
responsible for the CBR anisotropies, and
$\lambda$ is the scale of the density perturbation. This corresponds
to CBR temperature fluctuations
$(\Delta T/T)^2 \propto \theta^{(1-n)}$ for angles
$\theta > 30^\prime~\Omega_0^{1/2}$. The proposed observations
will be able to verify accurately the nearly 
scale invariant ``Harrison--Zel'dovich'' spectrum ($n=1$)
predicted by inflation. Any significant deviation from that
value would have extremely important consequences
for the inflationary paradigm. The COBE--DMR limit
on the spectral index after four years of observations
($n=1.1^{+0.2}_{-0.3}$, 68\% CL; \cite{Gorski96})
can be constrained $\sim 10$ times
better by the Planck results.

The proposed observations
will provide an additional, independent test for the inflationary model.
Temperature anisotropies on
large angular scales can be generated by gravitational
waves (tensor modes, $T$), in addition to the energy-density perturbation
component (scalar modes, $S$). Most inflationary models predict a
well determined, simple relation between
the ratio of these two components, $T/S$, and the spectral index $n$
\cite{Davis92}, \cite{Little92}: 
\begin{equation}
n \approx 1 - {1 \over 7}~{T \over S}.
\end{equation}
The Planck maps maybe able to verify this relationship,
since the temperature anisotropies from scalar and tensor modes
vary with multipoles in different ways.

A good satellite mission will be able not only to test the inflationary
concept but also to distinguish between various models
and determine inflationary parameters. There is an extensive
literature on what can be determined about inflation
such as the scalar and tensor power spectra, the energy scale
of inflation and so on
(see e.g. \cite{Steinhardt95}, \cite{Knox95}).
Such quality measurements lead also to good observations or constraints
for $\Omega_0$, $\Omega_{baryon}$, $\Lambda$, $H_0$, etc.
Sub--degree anisotropies are sensitive to the ionization
history of the universe. In fact, they can be erased
if the intergalactic medium underwent reionization
at high redshifts. Moreover, the temperature
anisotropies at small angular scales depend on other
key cosmological parameters, such as
the initial spectrum of irregularities, the baryon density
of the universe, the nature of dark matter, and the geometry of the universe
(see e.g. \cite{Crittenden93}, \cite{Kamionkowski94}, 
\cite{Hu94}, \cite{scott95}.
The Planck maps will provide constraints on these parameters
within the context of specific theoretical models.

Moreover, Planck should measure the Sunyaev--Zel'dovich effect for more 
than 1000 rich clusters, using the higher resolution bolometric channels. 
This will allow a rich analysis of clusters.
Combined with X--ray observations these
measurements can be used to estimate the Hubble constant $H_0$
as a second independent determination.

\subsubsection{Foreground Emissions}

In order to obtain these scientific goals, the measured
temperature fluctuations need to be well understood in terms
of the various components that add to the cosmological signal.
In fact, in addition
to the CBR temperature fluctuations, foreground structures will
be present
from weak, unresolved extragalactic sources and  from radiation of
Galactic origin (interstellar dust, free--free and
synchrotron radiation).
The Planck observations will reach the required control on
the foreground components in two ways.
First, the large sky coverage ($\geq$ 90\% of the sky)
will allow accurate modeling of these components where they
are dominant (e.g. Galactic radiation near the galactic plane).
Second, the observations will be performed
in a very broad spectral range. 

\begin{figure}
\epsfysize=3.5truein
\epsfverbosetrue
\centerline{\epsfbox[54 360 558 720]{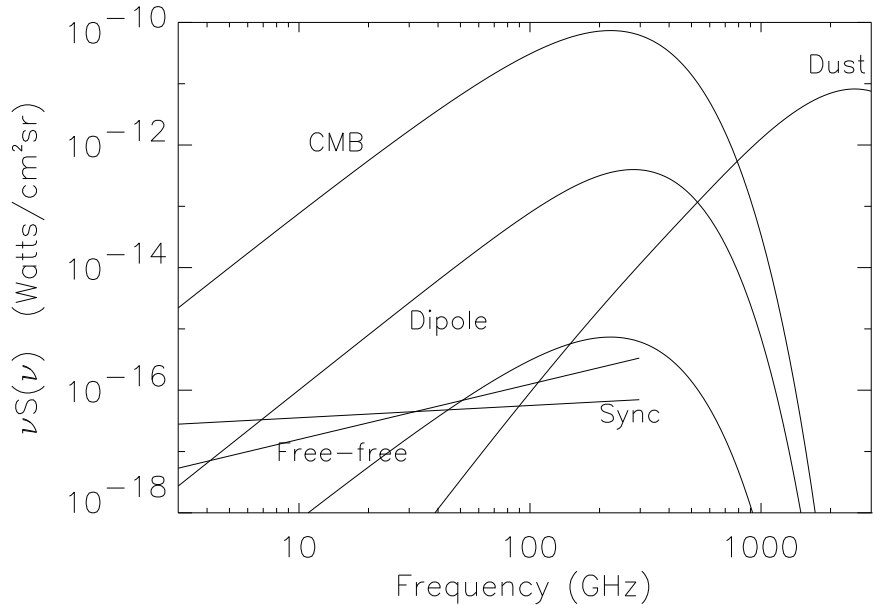}}
\caption{ 
The intensity of the microwave sky from 3 to 3000 GHz near a
Galactic latitude of $b=20^{\circ}$. 
The ordinate is the brightness of the sky times the frequency. 
which means the plot indicates the distribution of power per unit bandwidth. 
For synchrotron emission $S(\nu)\propto \nu^{-0.7}$; 
for free-free emission $S(\nu)\propto \nu^{-0.1}$; 
and for dust emission near 100 GHz $S(\nu)\propto \nu^{3.7}$. 
The scaling in effective temperature is $T(\nu )\propto\nu^{-2}S(\nu )$. 
The lowest Planck-like curve is for ${\rm T} = 27~\mu$K 
or anisotropy at the $10^{-5}$ level. }
\end{figure}

The Planck channels will span
the spectral region of minimum foreground intensity
(in the range 50--300 GHz), but with enough margin
at high and low frequency to monitor ``in real--time''
the effect of the various foreground components (see e.g. Brandt et al. 1994).
By using the Planck spectral information and modeling
the spectral dependence of galactic and extragalactic
emissions it will be possible to remove the foreground
contributions with high accuracy.

It should be noted that in most channels the ultimate limitation to the
cosmological information of high-quality CMB maps
is expected to be due to the residual uncertainties in the separation of
the foreground components rather than statistical noise. 
This explains why the overall design of Planck
is highly driven by the need of achieving as large a spectral coverage 
as possible.
Making the observations where the dominant foreground
components are different will permit a powerful cross check
on residual systematic errors in the CBR temperature fluctuation maps.

{\small
\begin{table}[t]
\begin{tabular}{|l||l|l|l|l|l|l|l|l|} 
\hline \hline
\multicolumn{9}{|c|}{\bf 1st Phase A Planck Payload Characteristics}\\
\hline
Telescope & \multicolumn{8}{c|}{1.5 m Diam. Gregorian; system emissivity $\leq$1\% }\\
          & \multicolumn{8}{c|}{Viewing direction offset $\geq 70^\circ$ from spin axis}\\
\hline
Instrument   & \multicolumn{4}{|c|}{LFI} & \multicolumn{4}{|c|}{HFI} \\
\hline
Center Frequency (GHz) &  31.5 &  53  &  90  &  125 & 140 &  222 & 400 & 714 \\
\hline
Wavelength (mm)		   &   9.5  &  5.7  &  3.3   &  2.4    &   2.1    &    1.4  &  0.75   & 0.42 \\
\hline
Bandwidth ($\Delta \nu \over \nu$) & 0.15 & 0.15 & 0.15 & 0.15& 0.4 & 0.5 & 0.7 & 0.6 \\
\hline
Detector Technology        & \multicolumn{4}{|c|}{HEMT receiver arrays} & \multicolumn{4}{|c|}{Bolometers arrays} \\
\hline
Detector Temperature       & \multicolumn{4}{c|}{$\sim$ 100 K }& \multicolumn{4}{c|}{0.1 - 0.15 K} \\
\hline
Cooling Requirements       & \multicolumn{4}{c|}{Passive}      & \multicolumn{4}{c|}{Cryocooler + Dilution system} \\
\hline
Number of Detectors        &   13   &   13   &   13    &   13  &  8  &  11 &  16  &  16 \\
\hline
Angular Resolution (arcmin)&   30   &   20  & 15 & 12 & 10.5 & 7.5 & 4.5 & 3 \\
\hline
Optical Efficiency & 1  & 1  & 1  & 1  & 0.3 & 0.3 & 0.3 & 0.3 \\
\hline
$\Delta T \over T$ Sensitivity (1$\sigma, 10^{-6}$ units,& 1.7 & 2.7 & 4.1 & 7.2 & 0.9&1.0&8.2&10$^4$\\
90\% sky coverage, 2 years) & & & & & & & & \\
\hline
$\Delta T \over T$ Sensitivity (1$\sigma, 10^{-6}$ units,& 0.6&0.9&1.4&2.4&0.3&0.3&2.7&5000 \\
2 \% sky coverage, 2 years) & & & & & & & & \\
\hline
\end{tabular}
\caption{Instrumental Parameters
for Planck the most important factors are the frequency coverage,
the angular resolution, sky coverage, and sensitivity.}
\label{tab:cobsam}
\end{table}
}
The Phase A Study provides a baseline design for the mission, spacecraft, and
instruments.  
Broad frequency coverage is achieved with arrays of HEMT amplifiers 
(30--100\,GHz) and bolometers (100--850\,GHz).  
The amplifiers use GaAs MMICs (monolithic microwave integrated circuits) 
cooled passively to about 100~K.
At the end of the Phase A study the number of detectors and focal plane
layout were optimized for the passively cooled LFI configuration.
This is summarized in Table \ref{tab:PlanckPA} 

{\small
\begin{table}[t]
\begin{tabular}{l|llll|lllll|} 
\multicolumn{10}{c}{ Final P{\sc hase} A I{\sc nstrument} S{\sc ummary} } \\
\hline \hline
Characteristic & \multicolumn{4}{|c|} {LFI} & \multicolumn{5}{|c|} {HFI} \\
\hline
Detector technology & \multicolumn{4}{|c|}{HEMT arrays} &
    \multicolumn{5}{|c|}{Bolometer arrays} \\
Detector temperature & \multicolumn{4}{|c|}{$\sim100$\,K} &
    \multicolumn{5}{|c|}{0.1--0.15\,K} \\
Center frequency [GHz]&31.5& 53 & 90 & 125 &  143 & 217 & 353 & 545 &857 \\
Number of detectors & 4 & 14 & 26 & 12 & 8 & 12 & 12 & 12 &12 \\
Angular resolution [$^\prime$] & 30 & 18 & 12 & 12 & 10.3 & 7.1 & 4.4 & 4.4 & 
4.4 \\
Bandwidth [$\Delta\nu/\nu$] & 0.15 & 0.15 & 0.15 & 0.15 & 0.37 & 0.37 & 0.37 & 
0.37 & 0.37 \\
Noise/res. element, & 21 & 20 & 39 & 97 & 3.3 & 5.5 & 33 & 210 & 11,000 \\
in 15\,months [$\mu$K] & & & & & & & & & \\ 
\hline
\end{tabular}
\caption{Proposed ``conservative'' Planck (COBRAS/SAMBA) instrument 
configuration at the end of the Phase A study. 
It was based upon MMIC GaAs and early micromesh bolometer technology
and a mix of cryocooler options. }
\label{tab:PlanckPA}
\end{table}
}

During the time since the Phase A study work has continued and it has
been learned that 
converting to InP HEMT amplifiers and actively cooling to 20~K provides
an inprovement in performance
and that changing some operating parameters the instrument might be
improved in other ways. 
Table \ref{tab:Planckmay} shows another possible configuration
taking advantage of the developments since the end of Phase A.

{\small
\begin{table}
\begin{tabular}{l|llll|llllll|} 
\multicolumn{11}{c}{{\bf P{\sc otential} P{\sc lanck} I{\sc nstrument} S{\sc ummary} }} \\
\hline \hline
Characteristic & \multicolumn{4}{|c|}{LFI} & \multicolumn{6}{|c|}{HFI} \\
\hline
Detector technology &\multicolumn{4}{|c|}{HEMT arrays} &
    \multicolumn{6}{|c|}{Bolometer arrays} \\
Detector temperature & \multicolumn{4}{|c|}{$20$\,K} &
    \multicolumn{6}{|c|}{0.1--0.15\,K} \\
Center frequency [GHz]& 30 & 44 & 70 & 100 &  100 & 143 & 217 & 353 & 545 & 857 \\
Number of detectors & 4 & 6 & 10 & 30 & 4 & 12 & 12 & 6 & 6 & 6 \\
Angular resolution [$^\prime$] & 34 & 23 & 16 & 10 & 14 & 10 & 7.1 &4.4 & 4.4 
& 4.4 \\
Bandwidth [$\Delta\nu/\nu$]& 0.20 & 0.20 &0.20&0.20&
     0.25&0.25&0.25&0.25&0.25&0.25 \\
Noise/res. element, & 5 & 7 & 10 & 17 & 3 & 4 & 7 & 41 & 240 & 10,600 \\
in 15\,months [$\mu$K] & & & & & & & & & & \\ 
Both polarizations? & yes & yes & yes & yes & no? & yes & yes & no? & no &no \\
\hline
\end{tabular}
\caption{Potential Planck detector configuration based upon active LFI cooling,
InP HEMT amplifiers, more advanced micromesh, filter, feedhorn, and polarizer
technology. }
\label{tab:Planckmay}
\end{table}
}


The need of accurate characterization of all non-cosmological
components, of course, brings the benefit of additional astrophysical 
information. The very large Planck data base, particularly when
combined with the IRAS
survey, can provide information on several
non--cosmological issues, such as the evolution
of starburst galaxies, the distribution of a cold--dust component,
or the study of low--mass star formation. 

\subsubsection{The Payload}

The Planck model payload consists mainly of a shielded,
off-axis Gregorian telescope, with a 
parabolic primary reflector and a secondary mirror, 
leading to an integrated instrument focal plane assembly.
The payload is part of a spinning spacecraft,
with a spin rate of 1 rpm.
The focal plane assembly is divided into low-frequency (LFI) 
and high-frequency (HFI) instrumentation according to the
technology of the detectors. Both the LFI and the HFI
are designed to produce high-sensitivity,
multifrequency measurements of the diffuse sky radiation.
The LFI will measure in four bands
in the frequency range 30--130 GHz (2.3--10 mm wavelength).
The HFI will measure in four channels in the range 140--800
GHz (0.4--2.1 mm wavelength). The highest frequency LFI channel
and the lowest HFI channel
overlap near the minimum foreground region.
Both the HFI and LFI teams are reinvestigating the optimal frequency bands.
Table \ref{tab:cobsam} summarizes the main characteristics
of the Planck payload. 

\subsubsection{The Main Optical System}

A clear field of view is necessary for the optics of a 
high-sensitivity CMB anisotropy experiment to avoid spurious signals arising
from the mirrors or from supports and mechanical mounting.
The off-axis Gregorian configuration has a
primary parabolic mirror of 1.5 meter,
and an elliptic secondary mirror (0.57 m diameter). 
Stray satellite radiation and other off--axis emissions are minimized
by underilluminating the low--emissivity optics.
The telescope
reimages the sky onto the focal plane instrument
located near the payload
platform. The telescope optical axis 
is offset by $70^\circ$ to $90^\circ$ from the spin axis. Thus 
at each spacecraft spin rotation the telescope pointing direction
sweeps a large (approaching a great) circle in the sky, according to the 
sky scan strategy.

Blockage is a particularly important factor
since several feeds and detectors are located in the focal plane,
and unwanted, local radiation (e.g. from the Earth, the Sun and the Moon)
needs to be efficiently rejected.
A large, flared shield surrounds the entire telescope and
focal plane assembly, to screen the detectors from contaminating
sources of radiation. The shield also plays an important role
as an element of the passive thermal control
of the spacecraft.

\begin{figure}
\centerline{\epsfxsize= 14 cm \epsfbox[0 0 493 246]{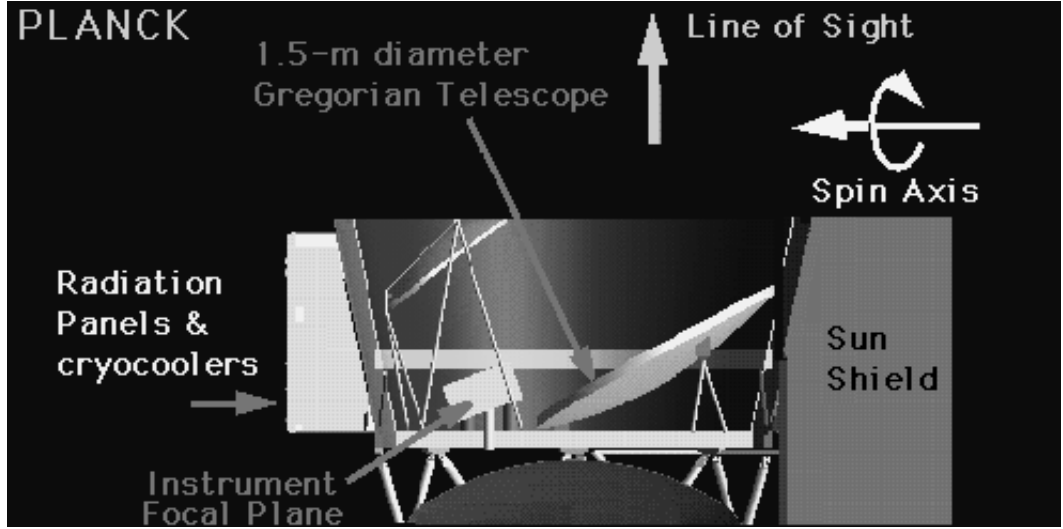}}
\caption
{Artist's concept of one possible configuration of the Planck Surveyor
optics and focal plane layout.
}
\label{PlanckC}
\end{figure}

\subsubsection{The Focal Plane Assembly}

The necessary wide spectral range requires the use of two
different technologies, bolometers and coherent receivers incorporated
in a single instrument. 
Both technologies have shown impressive progress in the last 
ten years or so, and more is expected in the near future.
The thermal requirements
of the two types of detectors are widely different.
The coherent radiometers (LFI), operating in the low frequency channels, 
give good performance at and operational temperature of $\sim 100$ K, which is 
achievable with passive cooling. 
Splitting the HEMT chain into cool and warmer portions leads
to a passive cooling temperature of 65~K and with active cooling 20~K.
The bolometers, on the other hand, require temperatures $\leq 0.15$~K
in order to reach their extraordinary sensitivity performances.
The main characteristics of the LFI and HFI are summarized
in Table \ref{tab:cobsam}.

The LFI consists of an array of 26 
corrugated, conical horns,
each  exploited in the two orthogonal polarization modes,
feeding a set of state--of--the--art,
high sensitivity receivers. The receivers
will be based on MMIC (Monolithic Microwave Integrated
Circuits) technology with HEMT (High Electron Mobility Transistor)
ultra--low noise amplifiers (see e.g. Pospieszalski et al. 1993). 
Since the whole LFI system will be passively cooled, it 
can be operated for a duration limited only by spacecraft consumables
(up to 5 years).
If actively cooled with a sorption cooler, it can still operate
for 5 years as there is no significant cryogen depeletion;
however, the HEMT chain must be broken into two sections.
The three lowest center frequencies of the LFI were chosen to match
the COBE-DMR channels, to facilitate the comparison of the product maps. 
The exact frequency bands are being reviewed in the upgraded design.

About 50 bolometers will be used in the HFI instrument,
which require cooling at $\sim 0.1$~K. The cooling
system combines active coolers reaching 4~K with a dilution
refrigeration system working at zero gravity. The refrigeration
system will include two pressurized tanks of $^3$He and $^4$He
for an operational lifetime of 2 years.

\subsubsection{Orbit and Sky Observation Strategy}

One of the main requirements for the Planck
mission is the need of a far--Earth orbit. This choice greatly reduces
the problem of unwanted radiation from the Earth 
which is a serious potential contaminant at the high goal sensitivity
and angular resolution.
The requirements on residual Earth radiation are
basically the same for the LFI and the HFI systems.
Adopting a low--earth orbit, such as that used by the COBE satellite,
the requirement on straylight and sidelobe rejection would be
a factor of $10^{13}$, which is beyond the 
capabilities of present microwave and sub--mm
systems and test equipment. 
Two orbits have been considered for Planck: 
a small orbit around the L5 Lagrangian point
of the Earth--Moon system, at a distance of about 400,000 km
from both the Earth and the Moon and the L2 Lagrange point of the Earth--Sun 
system. From the Earth--Moon Lagrange point the required rejection is relaxed 
by four orders of magnitude,
which is achievable with careful, standard optical designs.
For the Earth--Sun L2 point the situation for the Earth and Moon
is even better and the Sun is basically unchanged but because
the Earth, Moon, and Sun are all roughly in the same direction,
the spacecraft can be oriented very favorably.

These orbits are also very favorable 
from the point of view of passive cooling and thermal stability
\cite{Farquhar90}. 
The spacecraft will be normally operated in the anti-solar direction, 
with part of the sky observations performed within $\pm40^\circ$ from
anti--solar.

Other potential missions considered both a heliocentric orbit
and the Earth--Sun L2 point. All concerned seemed to have come to the
conclusion that the Earth--Sun L2 point is the best choice.
Operationally, it is difficult to find a more optimum location.

The main goal of the mission is to observe nearly the whole sky
($\gsim$ 90\%) with a sensitivity of 10--15 $\mu$K within the
two year mission lifetime. Deeper observation of a limited
($\sim 2\%$) sky region with low foregrounds 
could significantly contribute to the cosmological information. 
Simulations have shown that these observational objectives
can be achieved simultaneously in a natural way, using the spinning 
and orbit motion of the spacecraft,
with relatively simple schemes.

\section{Interpretation \& Future}

In five short years the field of CMB anisotropy observations and theory
has made great strides. Until April 1992 all plots of CMB anisotropy
showed only upper limits, except for the $\ell =1$ dipole.
Now we are beginning to trace out the shape of the power spectrum
and to make maps of the anisotropies.
This observational program promises to deliver a wealth of new information 
to cosmology and to connect it to other fields.
The COBE DMR has now released the full four-year data set.
We can expect little in the way of improvement compared to the final DMR results
from future experiments on the large angular scales
but scientific interest has moved to covering the full spectrum
and learning what the medium and small angular scales will tell us.
Already we are seeing plots showing the CMB anisotropy spectrum
related to and overlaid on the primordial density perturbation power spectrum
and attempts to reconstruct the inflaton potential.
These are the first steps in a new period of growth.

The last table gives an example of the level of sensitivity 
that might be achieved by the many experiments are underway, planned,
and approved.
Nearly every group has data 
under analysis and is also at work on developing new experiments.
Some of these are the natural extensions of the ongoing experiments.
Some groups are considering novel approaches.
Real long-term progress depends on avoiding the potential foregrounds: 
fluctuations of the atmosphere,  a source of noise that
largely overwhelms recent advances in detector technology,
and Galactic and extragalactic signals. 
This requires instruments having sufficient information
(usually only through multifrequency observations)
and observing frequencies to separate out the various components.
It also means going above the varying atmosphere. Collaborations
are working on long-duration ballooning instruments. Ultimately,
as COBE has shown, going to space really allows one to overcome the
atmospheric problem and to get data in a very stable and shielded
environment. The two selected satellite mission are actively being developed. 
We can anticipate a steady and significant advance in observations.
With the new data that are appearing, can be expected, and ultimately
will come from the Planck mission we can look forward to 
a very significant improvement in our knowledge of cosmology.

\begin{table*}[t]
\begin{center}
\begin{tabular}{|r||l|c|l|l|}
\hline
& 1997 & {\sc boom/max} & {\sc map}$^*$ & Planck$^*$ \\
\hline
$\Omega$                   & 0.01 - 2 & 6\%        & 18\%      & 1\%   \\
$\Omega_b$                 & 0.01$h^{-2}$ & 30\%       & 10\%      & 0.7\% \\
$\Lambda (\Omega_\Lambda)$ & $< 0.65$ & $\pm0.10$  & $\pm0.43$ & $\pm0.05$ \\
$\Omega_\nu$               &  $<2$    & $\pm0.25$  & $\pm0.08$ & $\pm0.03$ \\
$t_0 $                     & 12-18~Gyr &  ---       & ---       & ---       \\
$H_0$                      & 30-80 km/s/Mpc & 10\%       & 20\%      & 2\%   \\
$\sigma_8$                 & 0.5-0.6 & 30\%       & 30\%      & 10\%  \\
$Q$                        & $20 \pm 2\ \mu{\rm K}^*$ & $''$       & $''$      &$''$\\
$n_s$                      &  $1.0 \pm 0.5$ & 30\%       & 5\%       & 1\%   \\
$\tau$                     & 0.01 - 1 & $\pm0.5$   & $\pm0.2$  & $\pm0.15$ \\
$T_0$                      & $2.73 \pm 0.01^*$ &  ---       & ---       & ---       \\
$Y$                        & 0.2-0.25 & 10\%       & 10\%      & 7\%   \\
$T/S$                      & 0.0 - 1 & $\pm1.6$   & $\pm0.38$ & $\pm0.09$ \\
\hline
\end{tabular}
\caption{ Projected Parameter Errors:
Assumes variation around Standard CDM. ($^*$Bond et al., 1997.)
Note that parameters are not all independent, 
e.g., $H_0t_0=f(\Omega,\Lambda)$.
}
\end{center}
\end{table*}

\section*{Acknowledgements}
This work was supported in part
by the Director, Office of Energy Research, Office of High Energy and
Nuclear Physics, Division of High Energy Physics of he U.S. Department
of Energy under contract No. DE-AC03-76SF00098.



\begin{thebibliography}{}
\bibitem{ACE} ACE and BEAST. These two new projects are aimed at
using HEMTs between 26 and 100 GHz on both super-pressure
and conventional long-duration balloon platforms. The finest
angular resolution will be near $1/5^{\circ}$.
\bibitem{BOOMERanG}BOOMERanG is a collaboration between the Caltech,
Berkeley, Santa Barbara (Ruhl) and Rome groups.  It will use
bolometers to measure the anisotropy in the CMB between 90 and
410 GHz. The ultimate goal is a circumpolar Antarctic flight.
\bibitem{CBI}CBI: Cosmic Background Imager. This is an interferometer
that plans to produce maps of the microwave sky near 30 GHz.
\bibitem{MAT}MAT. This is the Mobile Anisotropy Telescope. It is
similar to QMAP but is designed to operate from the ground
in Chile.
\bibitem{OVRO}OVRO. The Owen's Valley Radio Observatoty telescopes
operate with various receivers between
15 and 30 GHz. The 40 m dish has a $2^{\prime}$ beam, and the
5.5 m has a $7.3^{\prime}$ beam. The experiments are aimed
primarily at small angular scales.
\bibitem{QMAP}QMAP. This is a balloon-borne experiment that uses a
combination of HEMTs and SIS detectors. The angular resolution is
$1/5^{\circ}$. This experiment is designed to produce
``true'' maps of the sky.
\bibitem{SASK}SASK. These experiments are based on HEMT amplifiers
operating between 26 and 46 GHz. They were performed in
Saskatoon, Saskatchewan CA. Three years of observations have gone into
the final data set.
\bibitem{SuZie}SuZIE is a bolometer-based experiment that
observes from the ground. It is primarily intended to measure the
SZ effect at high frequencies though it will also give information
on the anisotropy at small scales.
\bibitem{VCA}VCA: Very Compact Array. This is an interferometer
being developed at the University of Chicago. It will produce
maps of the CMB at 30 GHz and be sensitive to larger angular
scales than CBI.
\end{thebibliography}
\end{document}